\PassOptionsToPackage{svgnames, x11names, dvipsnames}{xcolor}
\newif\ifannotatedbuild
\ifdefined\annotatedbuild
  \annotatedbuildtrue
\else
  \annotatedbuildfalse
\fi

\ifannotatedbuild
  \documentclass[twocolumn,superscriptaddress, linenumbers]{revtex4-1}
\else
  \documentclass[twocolumn,superscriptaddress]{revtex4-1}
\fi
\usepackage{graphicx} 
\usepackage{xcolor}
\usepackage{amssymb}
\usepackage{standalone} \standaloneconfig{mode=buildnew} \usepackage{tikz}
\usepackage{xr-hyper}
\usepackage{amsmath} \usepackage[pdftex, hidelinks, bookmarks=true]{hyperref}
\usetikzlibrary{calc, positioning, arrows.meta} 
\usepackage{upgreek, textgreek}
\usepackage{booktabs}
\usepackage{float}
\usepackage[LGR,T1]{fontenc}
\usepackage[utf8x]{inputenc}
\usepackage{textalpha}
\usepackage{ulem}
\usepackage{marginnote}
\usepackage{tabularray}

\newcommand{\GCGC}{ {}^{lm} _{LM} L_R = 0 }
\ifannotatedbuild
  \newcommand{\MFnew}[1]{\textcolor{purple}{#1}}
\else
  \newcommand{\MFnew}[1]{#1}
  \newenvironment{linenumbers}{}{}
\fi

\begin{document}

\title{Data-efficient machine-learning of complex Fe--Mo intermetallics using domain knowledge of chemistry and crystallography}

\begin{abstract} 
  \begin{linenumbers}
      \section*{Abstract}
Atomistic simulations of multi-component systems require accurate descriptions of interatomic interactions to resolve energy differences between competing phases.
Particularly challenging are topologically close-packed (TCP) phases with structural similarities and nearly-degenerate different site occupations even in binary systems like Fe--Mo.  
In this work, data-efficient machine-learning (ML) models are presented that address this challenge by using features with domain knowledge of chemistry and crystallography, enabling accurate and robust predictions for the complex TCP phases \textit{R}, \textit{M}, \textit{P}, and $\delta$ with 11--14 WS after training on simple TCP phases \textit{A}15, $\sigma$, $\chi$, $\mu$, \textit{C}14, \textit{C}15, and \textit{C}36 with 2-5 \MFnew{Wyckoff sites (WS)}.
Several ML models based on kernel-ridge regression, multilayer perceptrons, and random forests are trained on fewer than 300 DFT calculations for the simple TCP phases in the Fe--Mo system.
Model performance is shown to improve systematically with increasing use of domain knowledge, reaching uncertainties below 25\thinspace meV/atom for the predicted convex hulls of the complex TCP phases and showing excellent agreement with DFT verification.
Complementary X-ray diffraction experiments and Rietveld analysis are conducted for a Fe--Mo \textit{R}-phase sample. 
The measured WS occupancies show excellent agreement with ML-model predictions obtained using the Bragg-Williams approximation at the same temperature.
  \end{linenumbers}
\end{abstract}

\author{Mariano Forti} \email{mariano.forti@rub.de}
\affiliation{Interdisciplinary Centre for Advanced Materials Simulation (ICAMS), Ruhr-Universit{\"a}t Bochum, 44801 Bochum, Germany}

\author{Alesya Malakhova} 
\affiliation{Interdisciplinary Centre for Advanced Materials Simulation (ICAMS), Ruhr-Universit{\"a}t Bochum, 44801 Bochum, Germany}

\author{Yury Lysogorskiy} \affiliation{Interdisciplinary Centre for Advanced Materials Simulation (ICAMS), Ruhr-Universit{\"a}t Bochum, 44801 Bochum, Germany}

\author{Wenhao Zhang}
\affiliation{CNRS-Saint-Gobain-NIMS, IRL 3629, Laboratory for Innovative Key Materials and Structures (LINK), 1-1 Namiki, Tsukuba, 305-0044, Ibaraki, Japan}
\affiliation{Research Center for Structural Materials, National Institute for Materials Science, 1-2-1 Sengen, Tsukuba, 305-0047, Ibaraki, Japan}

\author{Jean-Claude Crivello}
\affiliation{CNRS-Saint-Gobain-NIMS, IRL 3629, Laboratory for Innovative Key Materials and Structures (LINK), 1-1 Namiki, Tsukuba, 305-0044, Ibaraki, Japan}
\affiliation{Univ Paris Est Creteil, CNRS, ICMPE, UMR 7182, 2 rue Henri Dunant, 94320, Thiais, France}

\author{Jean-Marc Joubert}
\affiliation{Univ Paris Est Creteil, CNRS, ICMPE, UMR 7182, 2 rue Henri Dunant, 94320, Thiais, France}

\author{Ralf Drautz} \affiliation{Interdisciplinary Centre for Advanced Materials Simulation (ICAMS), Ruhr-Universit{\"a}t Bochum, 44801 Bochum, Germany}

\author{Thomas Hammerschmidt} 
\affiliation{Interdisciplinary Centre for Advanced Materials Simulation (ICAMS), Ruhr-Universit{\"a}t Bochum, 44801 Bochum, Germany}

\maketitle

\section*{Introduction}

Intermetallic phases are one of the most commonly observed groups of crystal structures.
The topologically close packed (TCP) phases are a subgroup of the intermetallic phases and known to form in numerous binary, ternary and multi-component alloys, see e.g. Refs.\cite{Sinha-1972, Joubert-08, Stein-21} for an overview.
They exhibit peculiar properties and play an important technological role as beneficial precipitates in light-weight steels ~\cite{wang_microstructure_2020,CHEN2024147062,sun_novel_2018,wan_kinetic_2018}, as detrimental precipitates in single-crystal superalloys ~\cite{Kossmann-15,Lopez-16,Wilson2017} and as potential crystal-structures for hydrogen storage, see e.g. ~Refs.~\cite{Joubert-01,Joubert-03,Merlino-16,Yartys-22}. 
In recent years, TCP phases have also been discussed in supramolecular chemistry where molecular self-assembly has been shown to form artificial molecular crystals with TCP structure with small molecules as building blocks instead of atoms, see e.g. Refs.~\cite{Montis-21,Zhuo-25}. 

\begin{figure}[H]
\centering
\includegraphics[width=\linewidth]{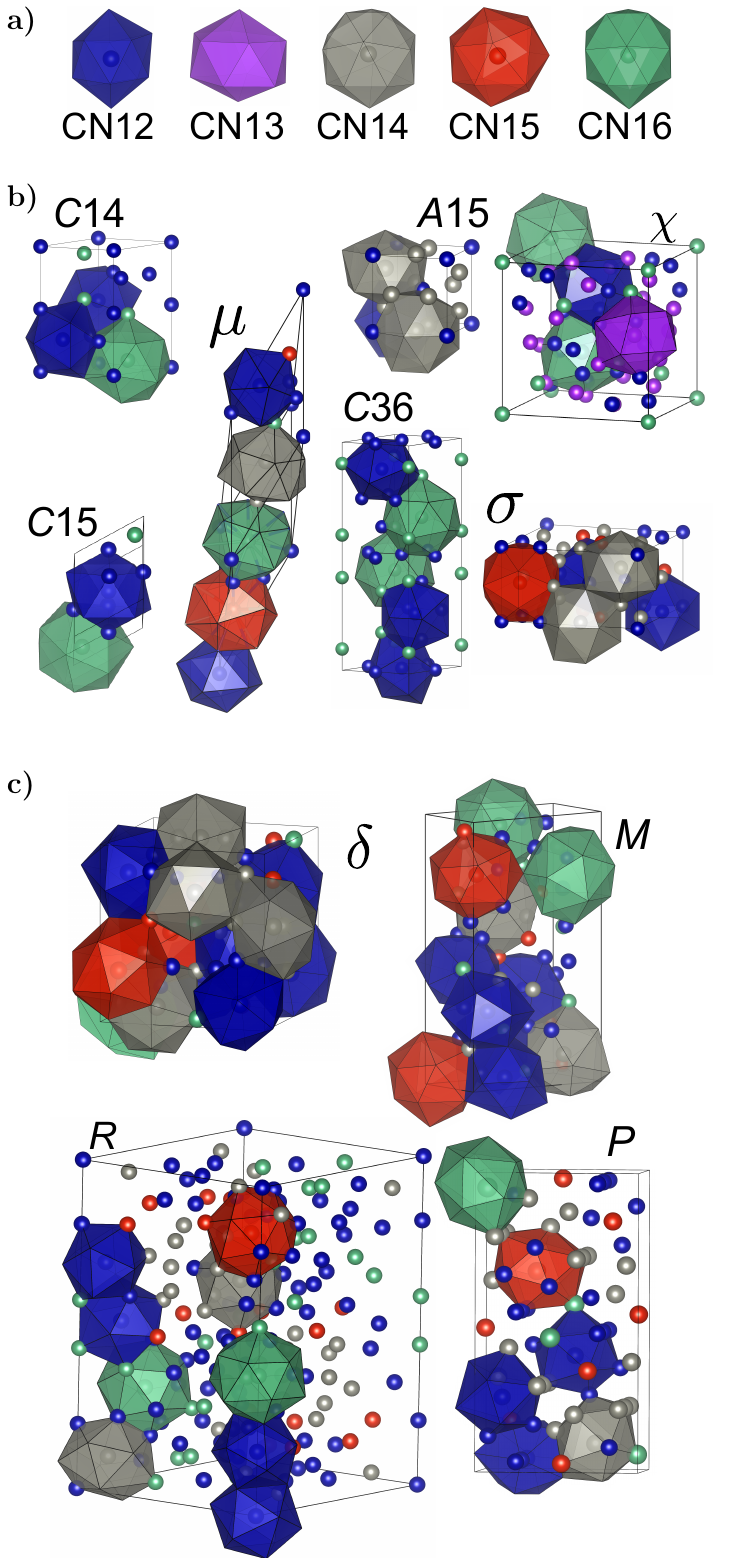}
\caption{\label{FIG_OVERVIEW_TCPS}
Overview of (a) FK polyhedrons with CN 12 to 16, (b) simple TCP phases with up to 5 WS in unit cells up to 30 atoms and (c) complex TCP phases with up to 14 WS in unit cells of up to 56 atoms.
The colors indicate the WS with the same CN.
The FK polyhedrons are shown for one atom per sublattice.
\MFnew{Crystal structures with coordination polyhedrons were generated with VESTA~\cite{Momma:ko5060}.}}
\end{figure}

The set of considered TCP phases contains \textit{A}15, $\mu$, $\sigma$, $\chi$, the Laves phases \textit{C}14, \textit{C}15, \textit{C}36, as well as the \textit{R}, \textit{M}, \textit{P}, and $\delta$ phase. 
Table \ref{TAB_TCP_WS} compiles prototypes and space groups (SG) of the TCP phases, as well as the details of the multiplicity and coordination number (CN) of their \MFnew{WS}. 
The crystal structures of these phases are geometrically closely related with only tetrahedral interstices, and in consequence lattice sites can have a limited number of coordination polyhedrons. 
Therefore, they can be seen as different periodic arrangements of the Frank-Kasper (FK) polyhedrons with nearest-neighbor coordination numbers (CN) from 12 to 16.
Each WS in the TCP phases corresponds to one of the FK polyhedrons and forms one of the sublattices. 
As illustrated in ~\autoref{FIG_OVERVIEW_TCPS} the TCP phases can be loosely grouped in the comparably simple TCP phases with 2~WS (\textit{A}15, \textit{C}15), 3 WS (\textit{C}14), 4 WS ($\chi$), and 5 WS (\textit{C}36, $\sigma$, $\mu$) in unit cells with up to 30 atoms, and the complex TCP phases with 11 WS (\textit{R}, \textit{M}), 12 WS (\textit{P}) and 14 WS ($\delta$) in unit cells with up to 56 atoms.
The major factors that govern the thermodynamic stability of the TCP phases are the average number of valence electrons, differences in atomic size of the constituent elements, and entropy contributions in some cases. 
Predictions of the thermodynamic stability have been performed by empirical maps of structural stability (e.g. Refs.~\cite{Seiser-11-1, Kossmann-15, Lopez-16}), by density-functional theory (DFT) calculations (e.g. Refs.\cite{berne_site_2001, Sluiter-03, Sluiter-07, Pavlu-10, Kabliman-12, Hammerschmidt-13}) and by CALPHAD approaches using experimental data and/or DFT calculations (e.g. Refs.~\cite{Fries-02, Dupin-06, Joubert-08, Granas-08, Pavlu-10, Crivello-10, Palumbo-11, Li-16, Ostrowska-2020, dosSantos-24}).
DFT calculations and atomistic approaches in general require the computation of the formation enthalpy of the possible permutations of all constituent elements on all WS.
This is also the case in thermodynamic modeling within the framework of the Compound Energy Formalism (CEF), which addresses the energy of multi-component and non-stoichiometric phases~\cite{Sundman-81}. 

In this approach, each WS is treated as a sublattice, and the distribution of atoms across these sites generates unique configurations—called end-members—for which the formation enthalpy must be expressed individually.
However, these permutations quickly lead to a combinatorial explosion which limits a computationally involved methods like DFT to the simpler TCP phases with binary or ternary composition while the complex TCP phases remain out of reach. 
It remains to be seen if computationally more efficient universal machine learning potentials (see e.g. Ref.\cite{riebesell_framework_2025} will reach sufficient accuracy in the future to overcome these limitations.

Here we present a machine learning (ML) approach to overcome these limitations and demonstrate a complete sampling of all TCP phases in a binary system with very high accuracy. 
The Fe--Mo system is chosen here as an example, as it is one of the few binary systems that forms simple TCP phases ($\sigma$, $\mu$, \textit{C}14) as well as a complex TCP phase (R). 
Nevertheless, our approach is generally applicable to all TCP phases in any compound.
The Fe--Mo phase diagram is well established on the basis of the experimentally observed phase equilibrium~\cite{Guillermet-82, Andersson-88, Eumann-08}. 
DFT calculations of formation enthalpies in Fe--Mo, however, are limited to simple TCP phases~\cite{Houserova-05, Ladines-15, alonso_stability_2011} while the complex TCP phases have not been accessible.
In this work, a machine-learning (ML) model is presented that allows accurate prediction of the structural stability of the complex TCP phases in the Fe--Mo system. 
The ML model is highly data-efficient and requires only the small set of simple TCP phases as training data.

\section*{Results}
\label{SECT_RESULTS}

\subsection*{DFT calculations for simple TCP phases}
\label{SECT_METHODS_DFT}

The training data for our ML models is generated by DFT calculations for the simple TCP phases in the Fe-Mo system. 
Details of the computational settings are compiled in the Methods section. 
For each simple TCP phase, we consider all possible permutations of Fe and Mo on all WS as given in ~\autoref{TAB_TCP_WS}.
This combinatorially complete sampling of chemical compositions leads to a total of 2$\times$(2\textsuperscript{2}+2\textsuperscript{2}+2\textsuperscript{3}+2\textsuperscript{4}+2\textsuperscript{5}+2\textsuperscript{5}+2\textsuperscript{5})=256 data points from the \textit{A}15, \textit{C}15, \textit{C}14, $\chi$, \textit{C}36, $\sigma$ and $\mu$ phase with a factor of two from fusion of magnetic and nonmagnetic DFT calculations.
This data set of the simple TCP phases was successively extended by few additional DFT calculations for complex TCP phases to validate the ML models developed in the following. 
A small subset of these calculations with sublattice occupations of the \textit{R} phase was included in the training data and is given Section S-I in the Supplementary Information.

\begin{table}[H]
  \caption{\label{TAB_TCP_WS} 
    Wyckoff sites~\cite{Sinha-1972,komura_crystal_1960} of the simple TCP phases \textit{A}15, \textit{C}15, \textit{C}14, \textit{C}36, $\sigma$, $\chi$, $\mu$, and the complex TCP phases \textit{R}, \textit{M}, \textit{P}, $\delta$. The CN13 polyhedron of $\chi$ is treated as distorted CN14.
}
\begin{tblr}{
  colspec = {Q[l, 0.31\linewidth] Q[c, 0.2\linewidth] Q[l, 0.47\linewidth]},
  colsep={1pt},
}
\hline \hline
Phase (prototype)  & SG  & WS (coordination number)\\
\hline \hline
\textit{A}15 (CrSi$_3$)   &$Pm\bar{3}n$  &  2$a$(12), 6$c$(14)     \\
\textit{C}15 (MgCu$_2$)   & $Fd\bar{3}m$ &  8$a$(16), 16$d$(12) \\
\textit{C}14 (MgZn$_2$)   &  $P6_3/mmc$ &  2$a$(12), 6$h$(12), 4$f$(16)\\
\textit{C}36 (MgNi$_2$)    & $P6_3/mmc$ &  4$e$(16), 4$f_1$(16), 4$f_2$(12), 6$g$(12), 6$h$(12)\\
$\sigma$  (CrFe/$\beta$-U)     & $P4_2/mnm$ &  2$a$(12), 4$f$(15), 8$i_1$(14), 8$i_2$(12), 8$j$(14)\\
$\chi$ ($\alpha-$Mn)  & $I4\bar{3}m$  &  2$a$(12), 8$c$(16),
             24$g$\textsubscript{D\textsubscript{1}}(12),
	         24$g$\textsubscript{D\textsubscript{2}}(13$\rightarrow$14)\\ 
$\mu$  (W$_6$Fe$_7$)         & $R\bar{3}m$  &  3$a$(12), 18$h$(12), 6$c_1$(15), 6$c_2$(16), 6$c_3$(14)                  \\
\hline
\textit{R} (Cr$_{16}$Mo$_{38}$Co$_{46}$)  & $R\bar{3}$ & 1$b_1$(12), 
            2$c_1$ (12), 
            6$f_1$ (12), 
            6$f_2$ (12), 
            6$f_3$ (12),  
            6$f_4$ (12), 
            6$f_5$ (14), 
            6$f_6$ (14), 
            6$f_7$ (15), 
            2$c_2$ (16),   
            6$f_8$ (16)          \\
\textit{M} (Nb$_{10}$Ni$_9$Al$_3$) & $Pnma$  &  4$c_1$   (12),   
            4$c_2$ (12), 
            4$c_3$ (12), 
            4$c_4$ (12), 
            4$c_5$ (12),
            4$c_6$ (14), 
            4$c_7$ (14), 
            4$c_8$ (15), 
            4$c_9$ (15), 
	        8$d_{10}$ (16), 
   	        8$d_{11}$ (16)  \\
\textit{P}  (Cr$_{18}$Mo$_{42}$Ni$_{40}$)  &  $Pnma$ &  4$c_1$  (12),
            4$c_2$  (12), 
            4$c_3$  (12), 
            4$c_4$  (14), 
            4$c_5$  (15), 
            4$c_6$  (16), 
            4$c_7$  (14), 
            4$c_8$  (12), 
            4$c_9$  (14), 
	        4$c_{10}$(15),    
	        8$d_{1}$ (12),    
            8$d_{2}$ (14) \\ 
$\delta$ (MoNi)      &  $P2_12_12_1$ &  4$a_1$(14),      4$a_2$ (14),             4$a_3$ (16), 
            4$a_4$ (12), 
            4$a_5$ (14), 
            4$a_6$ (12), 
            4$a_7$ (15), 
            4$a_8$ (12), 
            4$a_9$ (12), 
            4$a_{10}$ (12), 
            4$a_{11}$ (14), 
            4$a_{12}$ (12), 
            4$a_{13}$ (14), 
            4$a_{14}$ (15) \\ 
\hline
\hline
\end{tblr}
\end{table}

\begin{figure}[t] 
    \centering
    \includegraphics[width=0.95\linewidth, trim = 2cm 0.5 1.8cm 0.8cm, clip]{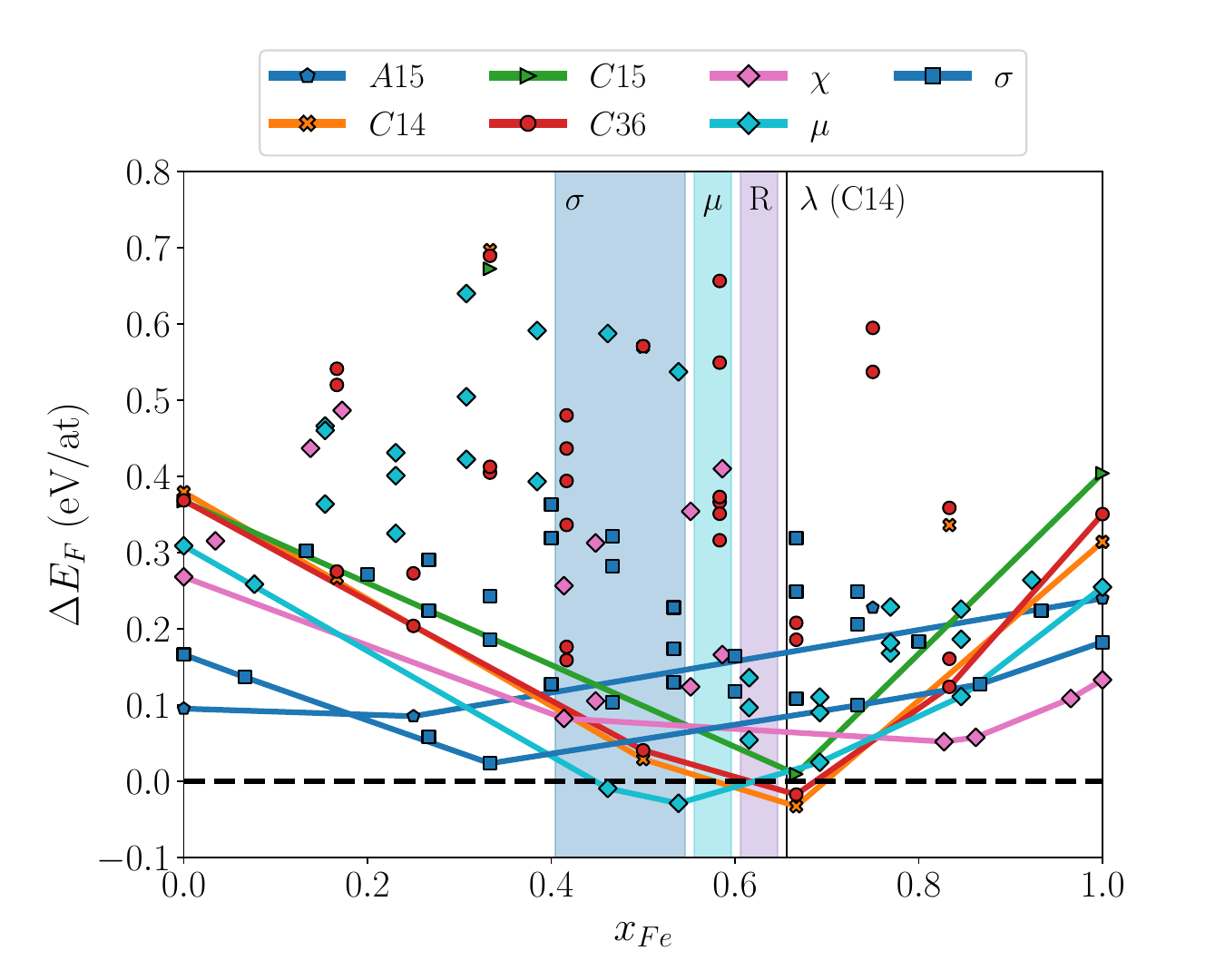}
\caption{\label{FIG_OVERVIEW_DFT}
DFT calculations of the formation enthalpy \MFnew{(eV/at)} of the simple TCP phases across the Fe--Mo composition range.
The vertical color bars indicate the experimentally observed stability ranges of the $\sigma$, $\mu$ and \textit{R} phases at finite temperatures.
The vertical black line indicates the \textit{C}14 Fe\textsubscript{2}Mo line compound denoted as $\lambda$ in the phase diagram.
Convex hull calculated from non-spin polarized calculations.}
\end{figure}

The spin-polarized DFT results for the simple TCP phases are in agreement with previous works~\cite{Houserova-05, Ladines-15} and in line with the Fe--Mo phase diagram. 
The non-spin-polarized DFT results compiled in \autoref{FIG_OVERVIEW_DFT} show the same sequence of structural stability. 
Both sets of DFT results consistently show negative formation energy of the $\mu$-phase near 55~at\% Fe in match with the experimental stability range of this phase at low temperatures. 
Also, the sequence of structural stability of the Fe\textsubscript{2}Mo Laves phases is \textit{C}14-\textit{C}36-\textit{C}15 in both cases, in line with the corresponding \textit{C}14 line compound in the phase diagram.
The $\sigma$-phase is unstable throughout the Fe--Mo composition range with positive formation energies of several tens of meV/atom in the experimentally observed range of stability in both sets of DFT calculations. 
We therefore expect that the high-temperature $\sigma$-phase is mostly stabilized by entropy contributions (as observed in comparable systems like Re--W~\cite{Crivello-10,Palumbo-14}) rather than by the influence of magnetism.
The formation energies of $\mu$ phase 55~at\% Fe and Laves phases at Fe$_2$Mo composition are slightly more negative in the non-spin-polarized DFT calculations. 

\subsection*{Machine learning with conventional features}
\label{SECT_CONVENTIONAL}

Our total data set for training with less than 300 data points is very small for ML standards and one would expect high errors in the predictions.
However, we show in the following that the utilization of domain knowledge in the construction of the descriptors leads to accurate predictions even for this very small set of training data.
Previous works~\cite{Seiser-11-1, Seiser-11-2, Bialon-16} showed the importance of atomic volumes, valence electrons and electronegativity as features to  qualitatively predict the likelihood for a given chemical composition to form a particular TCP phase. 
We take such conventional features as base line and represent chemical compositions of a particular TCP phase by sublattice occupations, e.g. 
$\sigma$-Fe\textsubscript{14}Mo\textsubscript{16} with Fe/Fe/Fe/Mo/Mo occupation of the 2$a$/4$f$/8$i_1$/8$i_2$/8$j$ sublattices. 
For each atom, we determine a variety of features using the \texttt{Matminer} package~\cite{ward_matminer_2018, ward_agrawal_choudary_wolverton_2016}. 

\begin{figure*}
  \centering
  \includegraphics[width=0.95\linewidth, trim=0 0.5cm 0 0.5cm, clip]{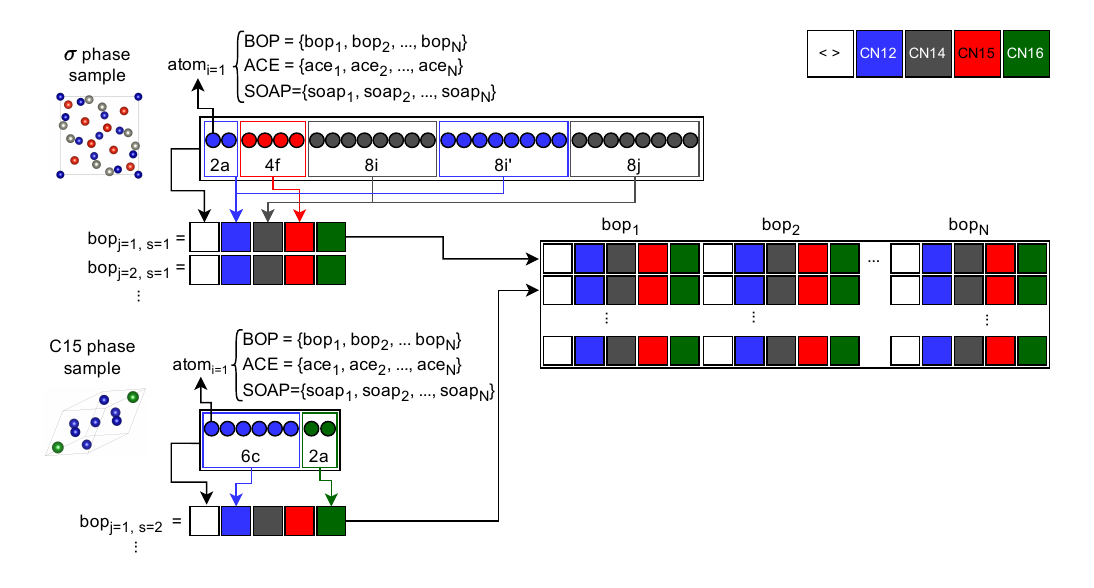}
  \caption{\label{FIG_DESCRIPTOR_ENGINEERING}
Feature-matrix construction shown for the $\sigma$ and \textit{C}15 phases with 30 and 8 atoms, respectively.
The features for each atom are computed with BOP, SOAP or ACE, and averaged over all atoms in the unit cell (UCavg) and over all sublattices with the same coordination number (CNavg), see \autoref{TAB_TCP_WS}.
The resulting row vectors with common length for all structures are stacked to form the feature matrix of the data set.}
\end{figure*}

\subsection*{Numerical representation of domain knowledge}\label{PAR:NumericalRepDK}

Domain knowledge on chemistry is introduced in a straight-forward way using the atomic properties of the chemical elements.  
In order to go beyond the conventional features described above which are agnostic of the crystal-lattice geometry, we simultaneously introduce domain knowledge of the crystallography of the TCP phases on the basis of the particular sets of WS given in~\autoref{TAB_TCP_WS}. 
Here, we implement a \textit{sequential encoding} of conventional atomic properties by WS. 
The sequential encoding fills a feature vector for each structure with atomic features of the atom in a site along fixed ordering of WS by coordination number. 
The united set of inequivalent sites of all TCP phases sets the common vector length.

In all following cases in this work, the resulting features depend in a non-linear way on the volume of the unit cells. 
Therefore, for a particular chemical composition and sublattice occupancy, instead of using generic crystal-structure prototypes with fixed unit cell and atomic positions, we compute the features for unit cells scaled to a volume estimated from a linear interpolation of DFT equilibrium volumes of the bcc ground-state of Fe and Mo following Vegard's law~\cite{vegard_konstitution_1921,denton_vegards_1991}. 
Section S-II in the Supplementary Information shows that this simple estimate captures the trend in atomic volumes for the simple TCP phases. 
Alternatively, one could use an additional ML layer to predict the equilibrium volume or atomic positions as shown for a data set of chemically complex $\sigma$-phases~\cite{crivello_supervised_2022}.
This pre-processing step with the nominal chemical composition is a natural way to equip the features with a certain degree of non-linearity to improve the ML performance. 

Chemistry-aware representations of the local atomic environments are put in place by per-atom features borrowed from different classes of interatomic potentials: smooth overlap of atomic positions (SOAP)~\cite{bartok_representing_2013}, atomic cluster expansions (ACE)~\cite{drautz_atomic_2019}, and bond-order potentials (BOP)~\cite{r_drautz_valence-dependent_2006, drautz_bond-order_2015}. 
Further details are described in the Methods section. 
A similar utilization of such descriptors of the local atomic environment from interatomic potentials was used recently for the unsupervised identification of crystal defects~\cite{Kyvala-2025}.
The SOAP approach is based on an atom-centered density that is represented in a basis of Gaussian radial functions and spherical harmonics. 
This leads to a local environment representation for each atom that picks up the density contributions from all neighboring atoms. 
The corresponding values of the evaluated basis functions are lexicographically ordered in our SOAP feature vector by lattice site, atomic species and order of the basis function. 
The construction of features with the ACE framework is also based on an atom-centered basis but exceeds the pair-wise summation of the SOAP approach with a systematic expansion in many-body summation terms. 
  Arranging the expansion coefficients \MFnew{corresponding to all the included body orders } lead to the ACE feature vector.
SOAP descriptors can be derived as a special case of ACE~\cite{drautz_atomic_2019}. In this work, they are treated separately and computed with different software packages to demonstrate this relationship also numerically in the trained ML models. 
\begin{figure*}
\centering \includegraphics[width=0.9\linewidth, trim=0 0.4cm 0 0.5cm, clip]{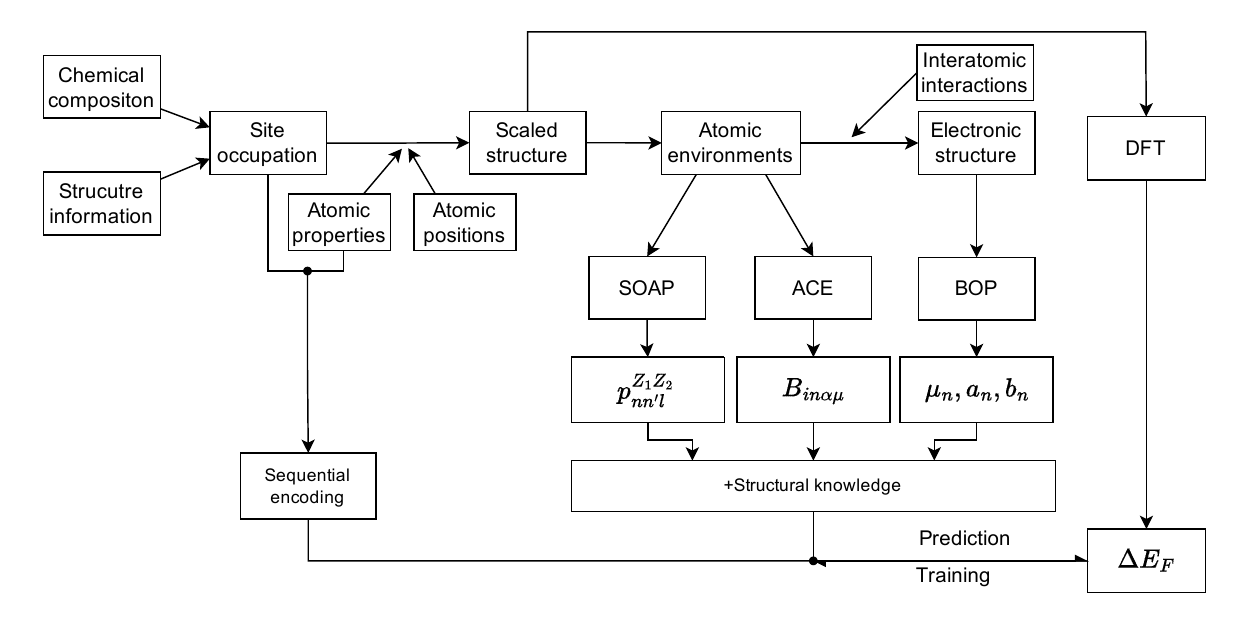}
  \caption{\label{Fig:GeneralMethodology}
Overview of ML workflow with numerical representation of domain knowledge on estimated compound-volume, coordination-resolved averaging and local chemistry.
Using this formalism, ML models are trained on the data set of simple TCP phases (\autoref{FIG_OVERVIEW_DFT}) in order to predict complex TCP phases (\autoref{Fig:prediction_for_M_d_P}).}
\end{figure*}
Complementary to the local geometry features of the SOAP and ACE formalism, we also consider local electronic-structure features in the framework of analytic BOPs. 
These features are based on the moments~\cite{cryot1967electronic, Ducastelle-70, Ducastelle-71} of the local, atom-resolved density-of-states (DOS) computed by recursion~\cite{Haydock-80-1} with a pairwise tight-binding Hamiltonian. 
Such electronic-structure features  computed with TB or DFT have been applied previously as representation of the local atomic environment, see e.g. Refs.~\cite{Turchi-83, Isayev-15, Hammerschmidt-16-2, sutton_crowd-sourcing_2019, naik_quantum-chemical_2023, Doesinger-25} and include a major part of the information already in the first few moments~\cite{Ducastelle-70, Bieber-83, Seiser-11-2,jenke_electronic_2018}.
ACE and SOAP features include categorical information on local chemical environment through a lexicographic ordering of the feature vector according to appearance of Fe and Mo atoms in the structure. 
The chemical information in the BOP features depends on the underlying pairwise Hamiltonians for specific pairs of interacting atoms. 
Here, on one hand, a chemistry-agnostic variant of BOP features is constructed using a canonical tight-binding model~\cite{andersen_electronic_1978}. 
On the other hand, a chemistry-aware variant with domain knowledge of the interatomic bond is constructed using projections of DFT eigenstates of specific pairs of atoms to a minimal basis~\cite{jenke_tight-binding_2021}. 

For the TCP phases, the particular challenge is to distinguish the distribution of chemical elements across the different WS that form a particular FK polyhedron. These distributions are compiled in Section S-III in the Supplementary Information for all TCP phases.
To achieve this, we introduce an averaging scheme that extends the conversion from per-atom features to common-size feature vectors. 
This is realized by computing not only the unit-cell average (UCavg) but also  coordination-resolved averages (CNavg) over the atoms of particular coordination numbers. 
All averages are taken for each per-atom feature and concatenated to a vector for each sublattice occupation of each TCP phase, see \autoref{FIG_DESCRIPTOR_ENGINEERING}. 
In this way, all TCP phases are represented by a common feature scheme. 
This approach to encode domain knowledge on crystallography follows previously introduced projections of feature vectors on a common feature basis (see e.g. Refs.~\cite{Rosenbrock-17}) and the common practice of merging sublattices with the same coordination number. 
This step is usually not needed if only one or two crystal structures are considered, see e.g. Refs.~\cite{crivello_supervised_2022, rao_machine_2022} while we train ML models for several TCP phases with different unit-cell sizes. 
Finally, a categorical feature for spin-polarization is merged into all feature sets to utilize both, spin-polarized and non-spin-polarized DFT calculations, as training data. 

\subsection*{Boost of ML performance by domain knowledge}

The different features, levels of domain knowledge, and regression algorithms, are independent of each other and result in the ML workflow in~\autoref{Fig:GeneralMethodology}.
For each crystal-structure prototype and chemical composition, the WS are occupied and conventional atomic features are computed. 
Domain knowledge on atomic volumes is applied to scale the structures according to their nominal chemical composition. Using these structures, the local atomic environments are encoded with chemistry-awareness using SOAP, ACE or BOP features. Finally, domain knowledge on the crystallography of the TCP phases is employed in terms of specific averaging schemes. 
The resulting feature sets are then used in supervised learning with kernel ridge regression (KRR), multi-layer perceptron (MLP) and random forest (RFR) as implemented in \texttt{scikit-learn}~\cite{scikit-learn}. Other regression schemes would require high data volumes for training, making them less suitable for our very small data set. 
The training efficiency is optimized by a greedy feature selection process with recursive forward feature addition described in the Methods section and in Section S-IV of the Supplementary Information. 
At the start of the ML model optimization, the SOAP, ACE, and BOP feature vectors contain 1650, 1800 and 410 scalar entries, respectively. 
The train/test splits of the data-set are carried out with constant relative contributions of the different TCP phases. 
All ML models are optimized using cross validation of the training split with diverse hyper-parameter grids.

\begin{figure}[H]
  \centering
    \includegraphics[width=0.9\linewidth,trim=0.35cm 0.3cm 0.05cm 0.2cm, clip]{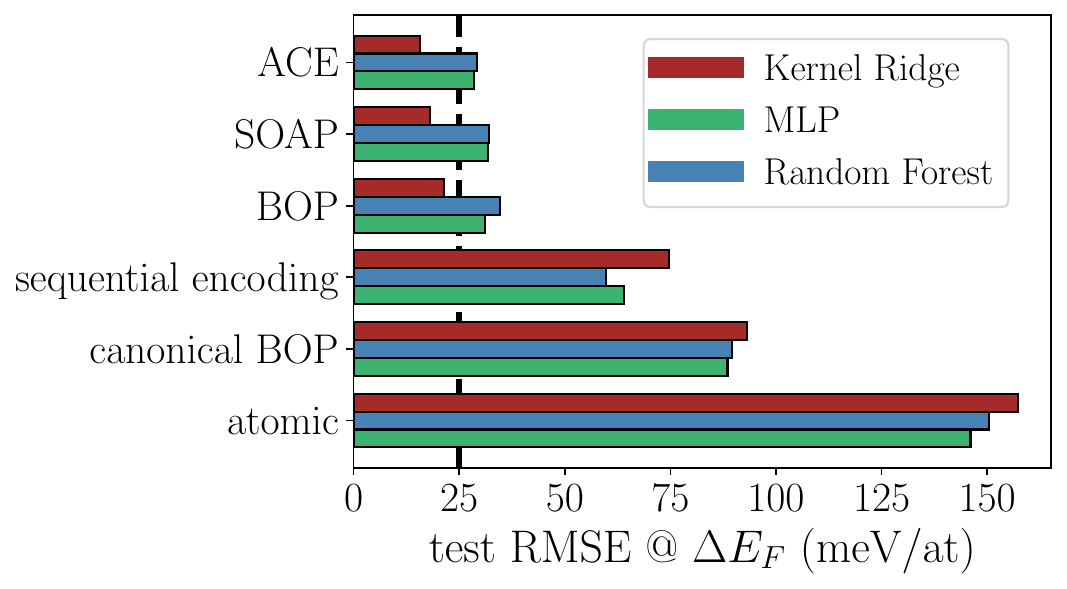}
  \caption{\label{Fig:comparison_feature_sets}
Performance of different feature/regression combinations for test split: conventional features (atomic), domain knowledge on chemistry-agnostic (canonical) BOP, of conventional features and chemistry-aware representations of the local atomic environment based on BOP, SOAP, and ACE.}
\end{figure}  

The performance of the resulting ML models is compared in~\autoref{Fig:comparison_feature_sets} regarding the root mean square errors (RMSE) on the \MFnew{
testing} set (The numerical values are given in Section S-V in the Supplementary Information).
Our baseline of conventional descriptors shows comparably large RMSEs of about 150\thinspace meV/atom for all three regression schemes. 
This is not unexpected given the variety of crystal structures and the ambiguity that different sublattice occupations can lead to the same chemical composition. 
E.g., the three sublattice occupations Fe/Fe/Fe/Mo/Mo, Fe/Fe/Mo/Fe/Mo, Fe/Fe/Mo/Mo/Fe lead to the same $\sigma$-Fe$_{14}$Mo$_{16}$ composition but have different DFT formation energies. 
The utilization of domain-knowledge on local atomic structure boosts the performance to about 90\thinspace meV/atom with the chemistry-agnostic variant of BOP features with approximate information of the local electronic structure. 
Higher levels of ML accuracy are reached by chemistry-aware domain-knowledge: Chemistry aware sequential encoding with ordering of conventional features according to WS and coordination polyhedra boosts the ML performance to about 60\thinspace meV/atom. 
With chemistry-specific domain knowledge of the local atomic environments, the performance reaches 25\thinspace meV/atom. 
The best performance is reached with utilizing all levels of domain-knowledge using KRR, with 20\thinspace meV/atom for ACE and 20-25\thinspace meV/atom for BOP and SOAP. 
This level of error is remarkably small given the small set of training data and the complexity of the crystal structures. 
The observed boosts by domain knowledge are fairly similar for the three considered regression schemes and demonstrates the robustness of the data efficiency of our approach.

\begin{figure}[H]
  \centering
  \includegraphics[width=\linewidth]{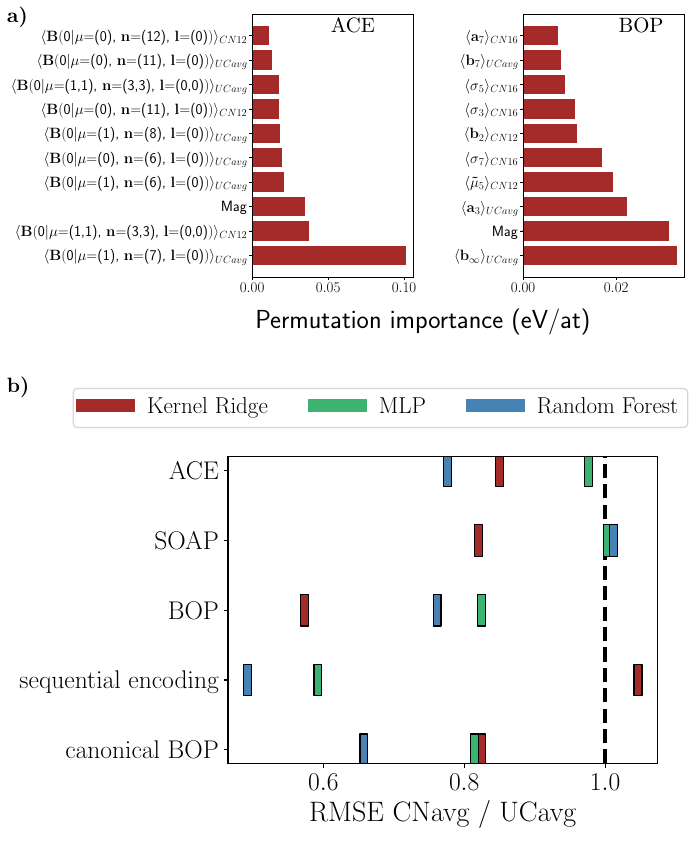}
  \caption{\label{Fig:Feature_importances}
Boost of model performance with domain knowledge in terms of (a) relative importance of particular features, illustrated for ACE and BOP.
For ACE features, the index of the expansion is indicated for a central atom (0) and its important interactions with the same atom ($\mu=0$) or the other species ($\mu=1$), while the expansion is symmetric with respect to the chemical element.
(b) coordination-resolved averages of per-atom features (CNavg) compared to averages over all atoms in the unit cell (UCavg).}
\end{figure}

The boosts in model performance can be directly attributed to the impact of particular features and averaging schemes shown in \autoref{Fig:Feature_importances}.
The reported ten most influential features from ACE and BOP representations with KRR are obtained with the permutation importance method~\cite{breiman_random_2001} that measures the importance of a descriptor as the change in model performance (RMSE on the test set) upon random permutation of its values.
Both, ACE and BOP, show high-importance of the categorical feature on the magnetic state with a similar importance of approximately \MFnew{30 meV/at}. 
Also, both models exhibit several features with averaging over WS with common~CN.
The dominant ACE features are primarily two-body terms associated with high-index radial basis functions, involving both mono-atomic and diatomic environments. 
In the BOP feature set, the most important descriptor is $\langle b_{\infty} \rangle_{UCavg}$, which is closely linked to the bandwidth of the DOS. 
\begin{figure*}
\centering
\includegraphics[width=\textwidth, trim=2.9cm 0.6cm 3.25cm 1.5cm, clip] {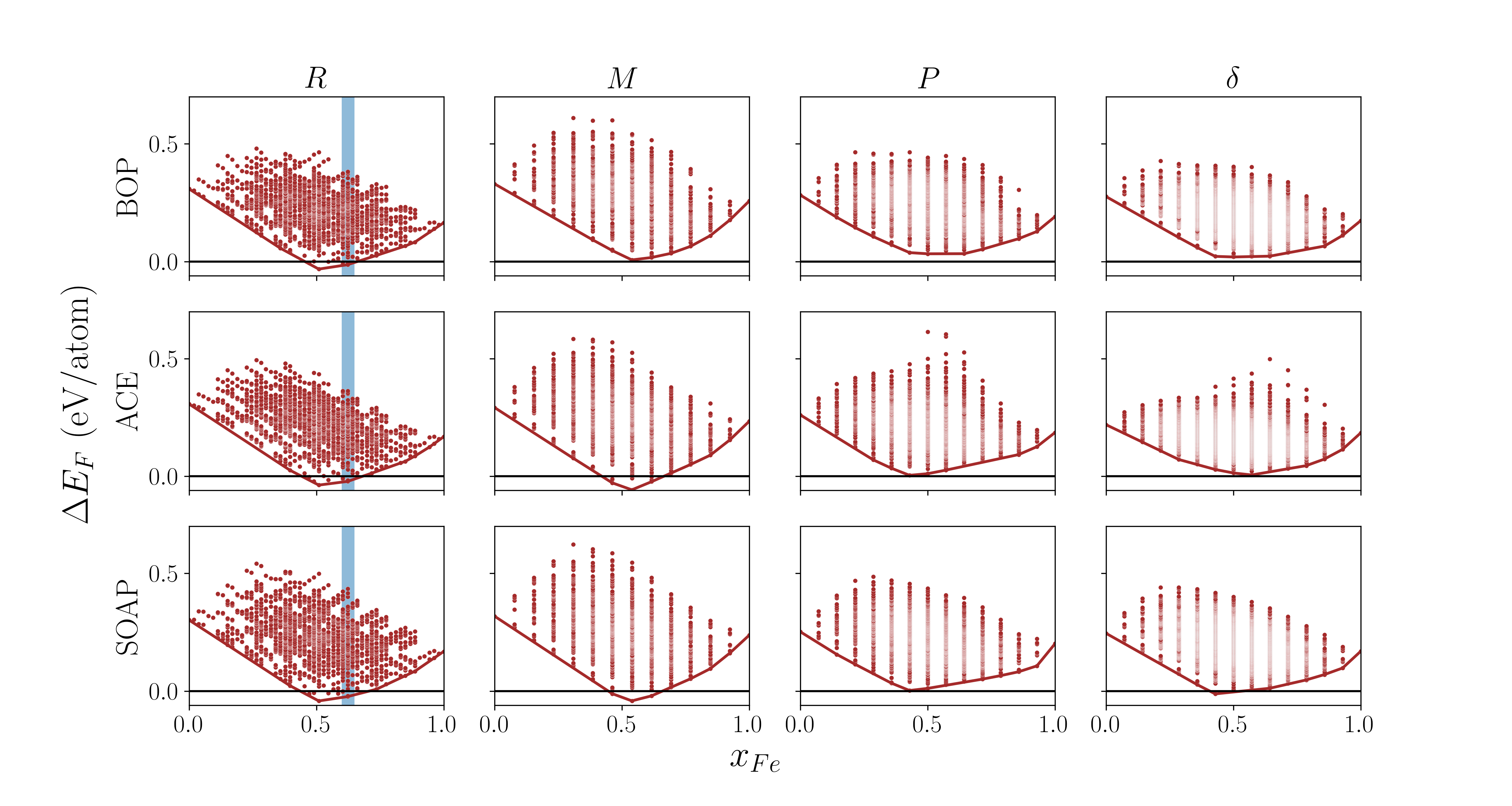}
  \caption{\label{Fig:prediction_for_M_d_P}
Formation enthalpy in \MFnew{eV/at}, and convex hull of complex TCP phases predicted with all levels of domain knowledge using SOAP, ACE and BOP descriptors, with models built from KRR in all cases.
The experimentally observed range of stability of the \textit{R} phase is indicated as blue bar in analogy to the overview of formation energies of the simple TCP phases in ~\autoref{FIG_OVERVIEW_DFT}.}
\end{figure*}
The high-index recursion coefficients and moments in the top-ranked BOP features correspond to longer paths with a more far-ranged sampling of geometry and chemistry. This reflects the need to distinguish not only the individual nearest-neighbor FK polyhedron of the TCP phases but also the arrangement of the different FK polyhedrons. 

The influence of the domain knowledge about the crystallography of the TCP phases is measured as the ratio of RMSE values obtained with CNavg over UCavg. The obtained values of nearly exclusively less than one show that nearly all combinations of features and regression schemes benefit considerably from the representation of domain knowledge on the crystal structure of TCP phases.

\subsection*{ML predictions for complex TCP phases}

With the generated ML models, we can for the first time predict the structural stability of the complex TCP phases \textit{R}, \textit{M}, \textit{P} and $\delta$ in a binary system with high accuracy and full sublattice complexity. 
The computed formation energies are shown in ~\autoref{Fig:prediction_for_M_d_P} for the three best models, i.e. with KRR and all levels of domain knowledge (\autoref{Fig:comparison_feature_sets}).  
In particular, we use volume-scaling according to chemical composition, coordination-resolved averaging of atomic features, and chemistry-aware representations of the local atomic environment from ACE, SOAP and BOP formalism. 
The range of chemical compositions is fully covered with a complete permutation of Fe and Mo on all sublattices following the procedure as for the training data on simple TCP phases.
This leads to $2^{11}+2^{11}+2^{12}+2^{14}=24576$ predicted data points for the \textit{R}, \textit{M}, \textit{P}, $\delta$ phase, respectively. 
The apparent patterns of the data points in ~\autoref{Fig:prediction_for_M_d_P} are due to the particular multiplicity of WS, see \autoref{TAB_TCP_WS}.
The results of the three ML models are overall very similar with only small differences for the \textit{M} and $\delta$ phase.

The \textit{R} and \textit{M} phases exhibit similar convex hulls across the whole composition range in all three models. 
The \textit{R} phase is consistently predicted to exhibit negative formation energies at around 1:1 composition and in the experimentally observed composition range. 
(The formation energies of all configurations on the convex hull by all ML models are given in Section S-VI in the Supplementary Information.)
Over this same range, the \textit{M} phase is predicted to be stable by ACE and SOAP and to be nearly stable by BOP.  
This close competition in structural stability of these two phases can be attributed to the crystallographic similarity of these two structures~\cite{Shoemaker-67}.
\begin{figure*}
\centering
 \includegraphics[width=\textwidth, trim=2.8cm 0.6cm 3.25cm 1.5cm,clip ] {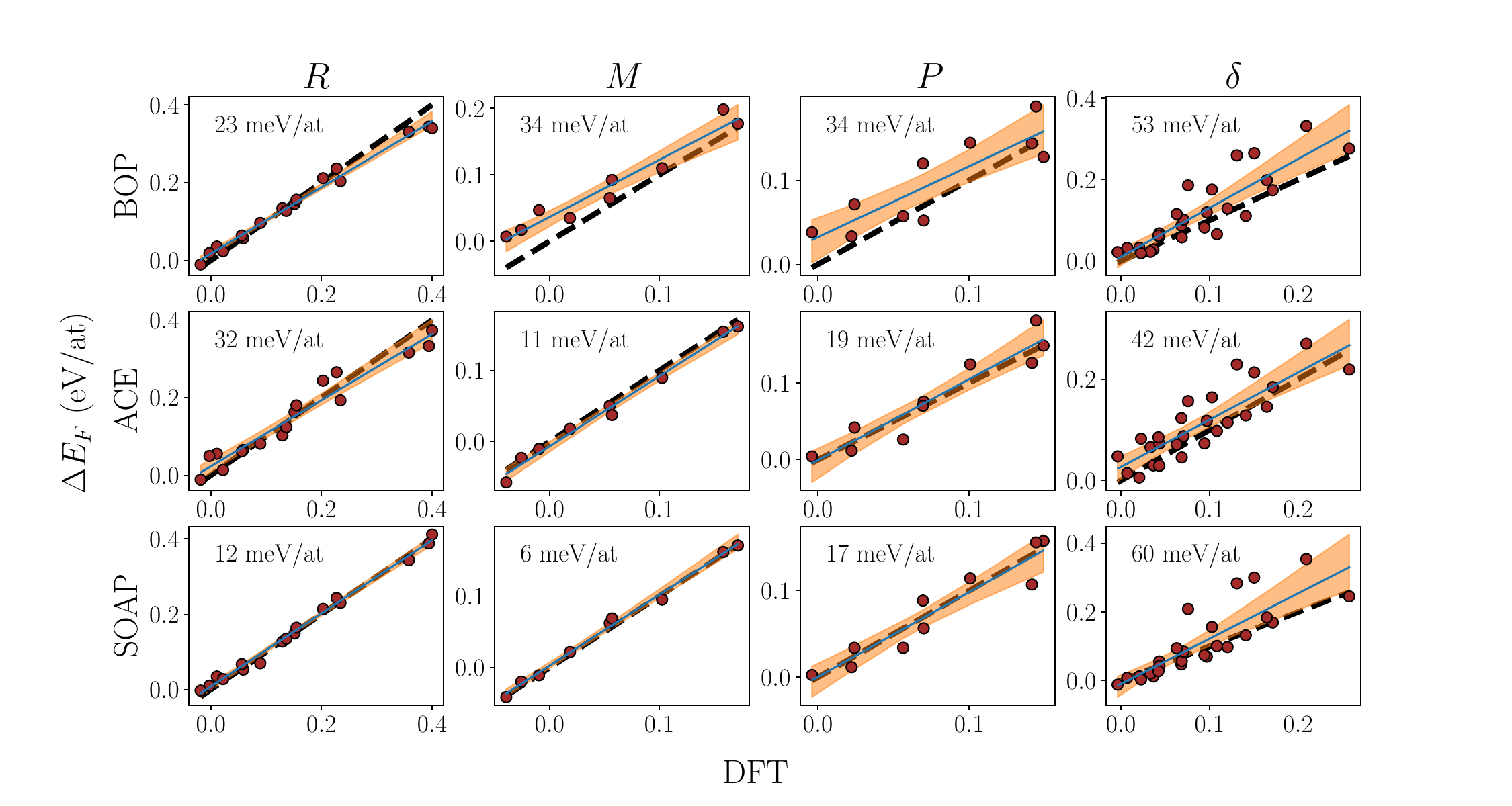}
  \caption{\label{Fig:Prediction_Validation}
Parity between ML predictions and DFT results of the formation enthalpy in \MFnew{eV/at}, for the validation set with confidence bands of the probability of predictions (orange) and dashed lines to indicate perfect agreement between ML models and DFT.
KRR is used.
\MFnew{For each prediction, the root mean squared error is indicated in the corresponding panel}.}
\end{figure*}
The allotropic transitions at the composition Fe\textsubscript{2}Mo from a $C$14 phase at low temperatures to a bcc/$mu$ two phase region at 1200 K and to an R phase at 1473 K is in line with the 
\MFnew{order in formation energies of the mentioned phases. 
For instance, The $C$14 phase at 0.66at\% Fe has $\Delta H_f$ = -0.032eV/at, the closest sample of the $\mu$ phase convex hull can be found at 0.53\%at Fe with a formation energy of -0.028eV/at, the closest R phase sample on the convex hull of that phase can be found at 0.63\%at Fe and has a formation energy of -0.023eV/at}.
    This indicates the importance of entropy \MFnew{and vibrational} contributions for the stabilization of the R \MFnew{and $\mu$} phases \MFnew{at high temperature}. 

A detailed analysis of the \textit{R} phase  configurations on the convex hull shows that increasing the Mo content fills the sublattices with Mo in the order CN16$\rightarrow$CN15$\rightarrow$CN14$\rightarrow$CN12. 
This progressive filling of FK polyhedra with larger coordination number by the larger Mo atoms is expected from the Kasper rules \cite{frank_complex_1958}. 

For the \textit{P} phase, the three models predict similar convex hulls with positive formation energies throughout. 
At around 45~at\% Fe, the \textit{P} phase takes a minimum that is apparently not sufficient to compete with the stable $\sigma$-phase in this composition range, see \autoref{FIG_OVERVIEW_DFT}. 
The $\delta$ phase is also predicted to be less stable than the \textit{R} and \textit{M} phase by all ML models, although with the comparably largest deviations of the convex hull. 
However, comparing the predicted minima of the convex hull of the $\delta$ phase with the other TCP phases, particularly \textit{C}14-Fe\textsubscript{2}Mo (see ~\autoref{FIG_OVERVIEW_DFT}), we do not expect that the $\delta$ phase is relevant even at elevated temperatures.

\subsection*{Validation of ML predictions}

The predictions of our ML models are validated with additional punctual DFT calculations for all structures that are predicted to be on the convex hulls by any of the ML models in~\autoref{Fig:prediction_for_M_d_P}. 
A comparison of all formation energies of the validation set obtained by DFT and by the three ML models for the \textit{R}, \textit{M}, \textit{P} and $\delta$ phases is compiled in~\autoref{Fig:Prediction_Validation}.
The predictions for the \textit{R} phase show error metrics similar to those obtained for the training data-set. 
For the \textit{P} and \textit{M} phases, the ML model constructed from BOP descriptors overestimates the formation energies slightly. For the \textit{M} phase, a small region of negative formation energies is observed in DFT calculations as well as in SOAP and ACE models, while only positive values are found with the ML model based on BOP descriptors. 
The $\delta$ phase appears to be challenging for all ML models with the largest errors for the BOP and SOAP features. Nevertheless, our DFT calculations confirm the instability of the $\delta$ phase predicted by all three ML models.
In fact, the $\delta-$phase is known to be stabilized in the Mo--Ni system, not in Fe--Mo~\cite{Shoemaker-1960}.

\begin{figure}
\centering
  \includegraphics[width=0.5\textwidth] {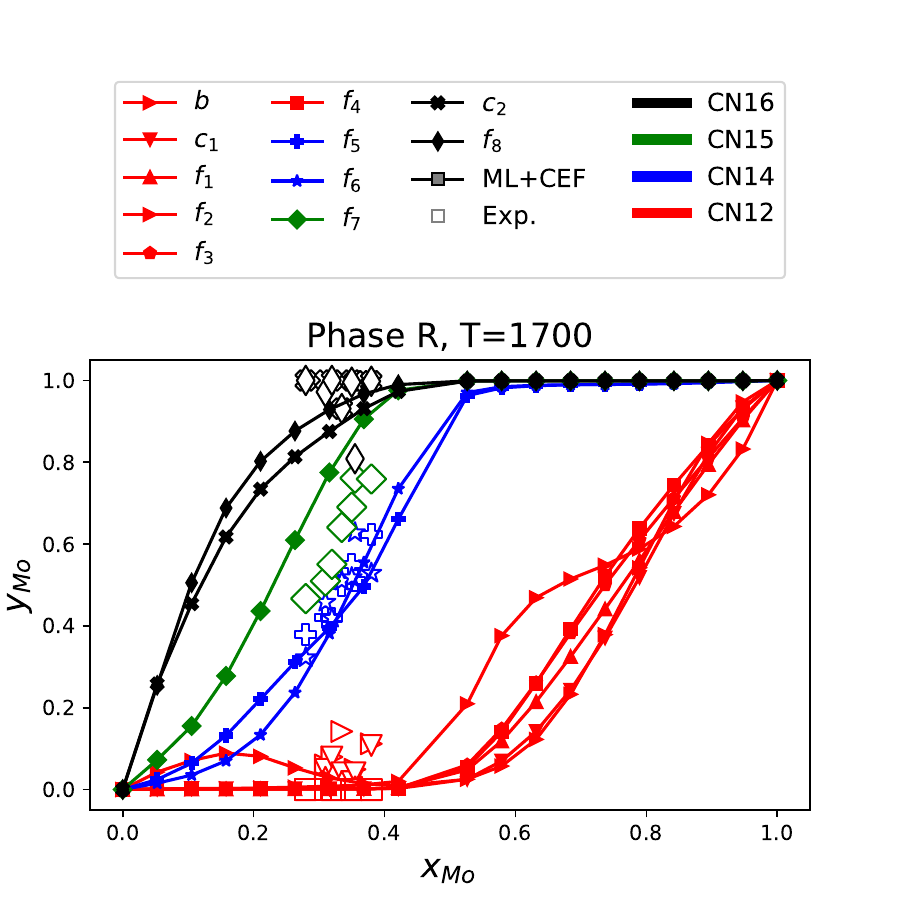}
  \caption{\label{Fig:SiteOccupancy}
\MFnew{Iron occupancy in the sublattices of Fe-Mo \textit{R} phase at 1700\thinspace K.
Empty symbols indicate measured occupancy, filled symbols show the solution of CEF problem from ML-predicted formation energies.
A color code is used to indicate the coordination number of the sublattices.}}
\end{figure}

As confirmation of the quality of the data generated by our model, we provide experimental data on the sublattice occupancy of the \textit{R} phase. The data is obtained by powder X-ray diffraction at a synchrotron source of a sample quenched from 1700\thinspace K and Rietveld refinement. Details on the experiment will be provided in an upcoming publication.
Parallel to these experiments, the sublattice occupations at the same composition and temperature were computed within the Bragg-William approximation~\cite{Fries-02, Dupin-06} using ML-predicted formation energies. 
The measured sublattice occupancy at the Fe--Mo composition of the experimental sample is compared to the computations of the sublattice occupancy across the full composition range in ~\autoref{Fig:SiteOccupancy}. 
The sublattice occupancy predicted at the chemical composition of the experimental sample is in excellent agreement with the measured data. 
In extension to the observation at \MFnew{\textit{T} = 0\thinspace K} from above, we find adherence to the Kasper rules also at high temperature in terms of the progressive filling of the sublattices by increasing CN for increasing Mo content at 1700\thinspace K.
As site occupancies are extremely sensitive to the ground state configurations and to the energy differences between these configurations, this can be considered as a strong validation of our ML approach.

\section*{Discussion}

In this work, we explore the importance of including domain knowledge in machine learning models for structural stability prediction. 
We demonstrate that highly predictive and data efficient ML models can be constructed with simple architectures and small training data-sets by utilizing domain knowledge. 
The case study is focused on the relative stability of intermetallic TCP phases in the Fe--Mo system. 
We generate DFT training data for the simple TCP phases \textit{A}15, $\sigma$, $\chi$, $\mu$, \textit{C}14, \textit{C}15, \textit{C}36 and construct ML models for the TCP phases \textit{R}, \textit{M}, \textit{P} and $\delta$ that would be combinatorially too complex for DFT calculations even in this binary compound. 

We construct features for our ML models that incorporate several kinds of domain knowledge. 
The numerical representation of the domain knowledge on chemistry is implemented as estimate of the equilibrium compound-volume. 
Domain knowledge on the crystallography of the TCP phases is implemented as descriptor averages across sets of lattice sites with the same coordination number. 
Chemistry-aware representations of local atomic environments are borrowed from interatomic potentials.
We demonstrate that increasing the level of domain knowledge systematically shifts the learning curves towards improved performance. 
This performance boost is independent of the regression algorithm and leads to significant improvements over commonly used descriptors of  geometry and composition encoding.

We present the first atomistic calculations of the ground-state convex hulls of the \textit{R}, \textit{M}, \textit{P} and $\delta$ phases with all sublattice occupations for a binary compound. 
The predictions of the best three ML models reach an RMSE on the test set of about 20\thinspace meV/atom. 
The validity of our ML models is demonstrated by additional punctual DFT calculations of the predicted convex hull. 

We obtain negative formation energy of the \textit{R} phase near compositions where this phase is also observed experimentally at high temperatures. 
The \textit{M} phase shows very similar structural stability and should be considered as competing phase in experimental studies of Fe--Mo. 
Our ML predictions indicate that the formation of \textit{P} and $\delta$ phases is not to be expected in the Fe--Mo system. 
Computed sublattice occupancies at 1700\thinspace K are in excellent agreement with according refined results from powder X-ray diffraction and confirm the Kasper rules of sublattice occupancy in the \textit{R} phase in the ground state and at high temperature.

\section*{Methods}

\subsection*{Density-functional theory calculations}\label{SUP:DFT_CALCULAITIONS}

The density functional theory (DFT) calculations are carried out with \texttt{VASP}~\cite{Hafner_doi:10.1002/jcc.21057} within a high-throughput environment \cite{Hammerschmidt-13}.
The general-gradient approximation~\cite{Perdew1996} and the projector augmented wave method~\cite{Blochl1994} are used.
With Brillouin zone integrations using Monkhost-Pack grids with densities of at least 0.02 / \AA\textsuperscript{3} and plane-wave expansions with cutoff energies of at least 400\thinspace meV, the difference in the total energies are converged within 0.1\thinspace meV/atom. 
We performed non-spin-polarized and spin-polarized DFT calculations, the latter with an initial ferromagnetic spin arrangement. 
All unit cells are fully relaxed with respect to internal degrees of freedom and cell shape until forces are less than 0.01 eV/ \AA.
The formation energies are determined by a workflow starting with full relaxation, followed by calculation of energy-volume curves around the relaxed structure and a fit of the energy-volume data to the Birch-Murnaghan equation of state~\cite{birch_finite_1947}. The  formation energies are taken relative to the respective ground state of Fe and Mo in bcc. 

\subsection*{Local atomic environment features}

SOAP features are calculated with the \texttt{Dscribe} package~\cite{himanen_dscribe_2020} using the feature vector element
\begin{equation}\label{Eqn:SOAPPowerSpectra}
  p_{nn'}^{Z_1, Z_2} = \pi\sqrt{\dfrac{8}{2l + 1}} \sum_m \left(c_{nlm}^{Z_1}\right)^{*}c_{n'lm}^{Z_2}
\end{equation}
\noindent with $c_{nlm}^Z$ being the projections of the atomic density $\rho^Z$ in a basis of spherical harmonics $Y_{lm}(\theta, \phi)$ and a Gaussian type radial functions $g_{nl}(r)$
\begin{equation} 
c^{Z}_{nlm} = \int \int \int dVg_{nl}(r)  Y_{lm}(\theta, \phi) \rho^Z(\vec{r}) .
\end{equation}
The radial components of the basis and the maximum angular part are considered up to $n \leq 5$ and $l\leq 4$, respectively.
The indices Z\textsubscript{1} and Z\textsubscript{2} encode the chemical species of the corresponding pair of atoms. The SOAP power spectra hence encodes chemistry information in the lexicographical order of the projections $p$ in the feature vector.\\

ACE features are computed in terms of the ACE basis functions using the \textit{python-ace} package~\cite{lysogorskiy_performant_2021} with the atom-resolved fingerprint of atom $i$ 
\begin{equation}\label{Eqn:AceBasisFunctions}
A_{i \mu \mathbf{nlm}} 
  =
  \prod_{t=1}^{\nu} \sum_{j} \delta_{\mu \mu_j} R_{nl}^{\mu \mu_j}(r_{ij}) Y_{lm}(\hat{\mathbf{r}}_{i, j})
\end{equation}
\noindent with chemical species $\mu$, and indexes of the radial and angular basis functions $\mathbf{nlm}$ for clusters of body order \MFnew{
$t \leq \nu$} . 
For the radial basis and the angular part we use Bessel functions with a maximal order of 4 and spherical harmonics with l $\leq$ 3 (s, p, and d), respectively, while we choose a maximal body order of 4.
The fingerprints $\mathbf{A}_{i \mu \mathbf{nlm}}$ are multiplied to generalized Clebsch-Gordan coefficients $\left( \GCGC \right)$ to obtain descriptors that are invariant under rotation and inversion~\cite{bochkarev_efficient_2022},
\begin{equation}\label{Eqn:ACERotInv}
    B_{i\mu\mathbf{nlL}} =
    \sum_{m} \Big( \GCGC \Big)
    A_{i \mu \mathbf{nlm}}
\end{equation}
\noindent where the summation is made over combinations of $\mathbf{m}$ that give non-zero contributions and the resulting expansion is expected to be rotationally invariant, so that its total orbital  moment $L_R$ is zero for the expansion.
Chemistry-awareness is enforced by constructing feature vectors where the basis functions are lexicographically ordered according to the chemical elements in the sites, the body order of the cluster and the radial functions and spherical harmonics. \\

BOP features are computed with the \texttt{BOPfox} software~\cite{hammerschmidt_bopfox_2019} on the basis of 
an orthogonal tight-binding model with pairwise Hamiltonians $H_{i\alpha j \beta}$ between atom $i$, orbital $\alpha$ and atom $j$, orbital $\beta$ in two-center approximation. 
The moments of the local electronic density of states of atom $i$, orbital $\alpha$ 

\begin{eqnarray}\label{Eqn:BOP_Moments}
  \mu^n _{i\alpha} &=& \langle i\alpha | H _{i\alpha} ^n |i\alpha \rangle\\
  &=& \sum_{j_k \beta_k}
  H_{i\alpha j_1 \beta_1} H_{j_1 \beta_1 j_2 \beta_2} \ldots H_{j_{n-1} \beta_{n-1} i\alpha}
\end{eqnarray}

and the corresponding recursion coefficients $a_{i\alpha}^{(n)}$ and $b_{i\alpha}^{(n)}$~\cite{Aoki-93-2} are determined with an iterative scheme~\cite{Haydock-80-2, Horsfield-96-3} based on the Lanczos algorithm~\cite{1950Lanczos}. 
For the chemistry-agnostic BOP variant, we use a canonical d-valent model~\cite{andersen_electronic_1978} as pairwise tight-binding Hamiltonian $H_{i\alpha j \beta}$. 
The chemistry-aware variant is based on pair-specific $H_{i\alpha j \beta}$ obtained by downfolding of DFT eigenstates to a minimal basis~\cite{jenke_tight-binding_2021}.

\subsection*{Machine learning}

Recursive feature selection with cross validation was used in all combinations of features and regression algorithms. 
The algorithm can be summarized as follows:

\begin{enumerate}
  \item obtain 
    \MFnew{a} test-train split on the data-set, \MFnew{stratified on the phases}.
  \item on the train split, perform stratified cross-validation grid search to fit \MFnew{a} regressor on each of the candidate scalar descriptors,
  \item identify \MFnew{a} scalar descriptor that results in the model with the best score on the test split,
  \item add it to \MFnew{a} selection vector and remove \MFnew{it} from the candidate vector,
  \item for each remaining scalar descriptors, perform another stratified cross-validation grid search using the descriptors in selection vector plus \MFnew{the} current component,
  \item add \MFnew{the} scalar descriptor that produces \MFnew{the} best performance on the test split to the selection vector and remove it from the vector of candidate descriptors,
  \item repeat 4 and 5 until selection vector is sufficiently long or until overfitting is observed, for example by \MFnew{sustained} decreasing performance of the fitted model.
\end{enumerate}

More details on the algorithm and detailed learning curves during feature selection are given in Section S-IV in the Supplementary Information.
 
For each regressor/feature combination, at least ten lists of selected features were constructed with sample shuffling in-between.
All collected selections of descriptors are used in ML pipelines as feature transformers. The pipelines are then used to build the voting regressor. This ensures  reproducibility of the predictions and well determined uncertainties. 

\section*{Data availability}

The DFT data and ML predictions will be made publicly available after acceptance of the manuscript.

\section*{Code availability}

\MFnew{
Analysis and ML training is based on python modules and jupyter notebooks. 
All plots are generated with the \texttt{matplotlib} package~\cite{Hunter:2007}.
Data tables are handled with the \texttt{pandas} package~\cite{mckinney-proc-scipy-2010}.
Convex hulls are calculated with the \texttt{Qhull} code~\cite{barber1996quickhull} as interfaced in the \texttt{scipy} package~\cite{2020SciPy-NMeth}.
}
The jupyter notebooks and the trained ML models will be made publicly available after acceptance of the manuscript.

\section*{Acknowledgments} 

This work is a part of the project ”Artificial Intelligence for Intermetallic Materials” - AIIM (ref. ANR-22-CE92-0036) funded and supported by the French National Research Agency (ANR) and the German Deutsche Forschungsgemeinschaft (DFG No. 505643559). 
We acknowledge Aparna Subramanyam and Jan Jenke for discussions in the initial stage of developing the computational workflow. 
We acknowledge A. Flores for the preparation of the sample and the synchrotron measurement and the team of the CRISTAL beamline at synchrotron SOLEIL. 
A part of DFT calculations were performed using HPC resources from GENCI-CINES (Grant A0060906175).
This preprint has not undergone peer review or any post-submission improvements or corrections. 
The Version of Record of this article is published in npj Computational materials, and is available online at https://doi.org/10.1038/s41524-026-02070-5.

\section*{Author contributions} 

TH conceived the project. 
MF, AM, YL and TH developed the ML workflow. 
WZ and JCC computed the site-occupancy. 
JMJ performed the experiments and the Rietveld refinement. 
\MFnew{TH and RD provided supervision and feedback during the realization of the project.}
MF and TH performed the DFT calculations and drafted the manuscript. 
Results were discussed with all authors. 
All authors contributed to the final version.

\section*{Competing Interests} The authors declare no competing financial or non-financial interests.

\subsection*{Correspondence} Correspondence and requests should be addressed to Mariano Forti.

\section*{References}
\bibliographystyle{naturemag}
\bibliography{main.bib}

\end{document}


\title{ Supplementary information supporting the article ``Data-efficient machine-learning of Fe--Mo complex intermetallics with domain knowledge of chemistry and crystallography''}



\author{Mariano Forti} \email{mariano.forti@icams.rub.de}
\affiliation{Interdisciplinary Centre for Advanced Materials Simulation (ICAMS), Ruhr-Universit{\"a}t Bochum, 44801 Bochum, Germany}

\author{Alesya Malakhova} 
\affiliation{Interdisciplinary Centre for Advanced Materials Simulation (ICAMS), Ruhr-Universit{\"a}t Bochum, 44801 Bochum, Germany}

\author{Yury Lysogorskiy} \affiliation{Interdisciplinary Centre for Advanced Materials Simulation (ICAMS), Ruhr-Universit{\"a}t Bochum, 44801 Bochum, Germany}

\author{Wenhao Zhang}
\affiliation{CNRS-Saint-Gobain-NIMS, IRL 3629, Laboratory for Innovative Key Materials and Structures (LINK), 1-1 Namiki, Tsukuba, 305-0044, Ibaraki, Japan}
\affiliation{Research Center for Structural Materials, National Institute for Materials Science, 1-2-1 Sengen, Tsukuba, 305-0047, Ibaraki, Japan}

\author{Jean-Claude Crivello}
\affiliation{CNRS-Saint-Gobain-NIMS, IRL 3629, Laboratory for Innovative Key Materials and Structures (LINK), 1-1 Namiki, Tsukuba, 305-0044, Ibaraki, Japan}
\affiliation{Univ Paris Est Creteil, CNRS, ICMPE, UMR 7182, 2 rue Henri Dunant, 94320, Thiais, France}

\author{Jean-Marc Joubert}
\affiliation{Univ Paris Est Creteil, CNRS, ICMPE, UMR 7182, 2 rue Henri Dunant, 94320, Thiais, France}

\author{Ralf Drautz} \affiliation{Interdisciplinary Centre for Advanced Materials Simulation (ICAMS), Ruhr-Universit{\"a}t Bochum, 44801 Bochum, Germany}

\author{Thomas Hammerschmidt} 
\affiliation{Interdisciplinary Centre for Advanced Materials Simulation (ICAMS), Ruhr-Universit{\"a}t Bochum, 44801 Bochum, Germany}

\maketitle

\section{DFT calculations for selected \textit{R}-phase configurations}\label{SUP:CleanUpTrainingSet} 

The complete list of R phase configurations in given in \autoref{TAB:R_in_TrainSet}.

\begingroup

\begin{center}
\tablecaption{\protect\label{TAB:R_in_TrainSet} 
    DFT calculations for selected \textit{R}-phase configurations.
    The strings in the `Configuration' column represent the occupation of the WS as indicated in the table \ref{TAB_TCP_WS}, where `A' stands for Fe and `B' for Mo.
  }
  \tablehead{
  \toprule[1.4pt]
 {}
 & ~~~~~Configuration~~~~~
               & ~Mag~
                    & ~$x_{Mo}$~~~
                            & ~~~$x_{Fe}$~~
                                    & ~$B_0$ (GPa)~
                                              & ~$V_0$ ($\AA^3/at$)~
                                                       ~& $\Delta E_F$ (eV/at)~ \\
}
\begin{supertabular}{lllccccc}
  \toprule
\multirow[t]{30}{*}
 & AAAAAAAAAAA & NM & 0.000 & 1.000 & 274 & 10.49 & 0.166 \\
\cline{2-8}
 & BAAAAAAAAAA & NM & 0.019 & 0.981 & 273 & 10.59 & 0.159 \\
\cline{2-8}
 & AAAAAAAAABA & NM & 0.038 & 0.962 & 274 & 10.67 & 0.142 \\
\cline{2-8}
 & BBAAAAAAAAA & NM & 0.057 & 0.943 & 271 & 10.80 & 0.163 \\
\cline{2-8}
 & AAAAAAAAAAB & NM & 0.113 & 0.887 & 277 & 11.00 & 0.082 \\
\cline{2-8}
 & BAAAAAABAAA & NM & 0.132 & 0.868 & 269 & 11.21 & 0.130 \\
\cline{2-8}
 & AAAAAAAAABB & NM & 0.151 & 0.849 & 278 & 11.16 & 0.063 \\
\cline{2-8}
 & ABAAAAAAABB & NM & 0.189 & 0.811 & 275 & 11.39 & 0.075 \\
\cline{2-8}
 & AAAABAAAABB & NM & 0.264 & 0.736 & 263 & 11.96 & 0.160 \\
\cline{2-8}
 & AAAAAAAABBB & NM & 0.264 & 0.736 & 276 & 11.75 & 0.029 \\
\cline{2-8}
 & ABAAAAAABBB & NM & 0.302 & 0.698 & 273 & 11.97 & 0.048 \\
\cline{2-8}
 & AAAAABAABBB & NM & 0.377 & 0.623 & 263 & 12.54 & 0.120 \\
\cline{2-8}
 & AAAAAABABBB & NM & 0.377 & 0.623 & 276 & 12.31 & -0.022 \\
\cline{2-8}
 & AAAAAAABBBB & NM & 0.377 & 0.623 & 276 & 12.31 & -0.023 \\
\cline{2-8}
 & AAAABABABAB & NM & 0.453 & 0.547 & 263 & 12.91 & 0.093 \\
\cline{2-8}
 & AAAABABABBB & NM & 0.491 & 0.509 & 265 & 13.06 & 0.065 \\
\cline{2-8}
 & AAAAABBABBB & NM & 0.491 & 0.509 & 265 & 13.09 & 0.067 \\
\cline{2-8}
 & AAAAAABBBBB & NM & 0.491 & 0.509 & 274 & 12.91 & -0.043 \\
\cline{2-8}
 & BBAAAABBBBB & NM & 0.547 & 0.453 & 268 & 13.30 & 0.017 \\
\cline{2-8}
 & AABABABABAB & NM & 0.566 & 0.434 & 254 & 13.66 & 0.166 \\
\cline{2-8}
 & AAABAABBBBB & NM & 0.604 & 0.396 & 265 & 13.66 & 0.023 \\
\cline{2-8}
 & AAAAABBBBBB & NM & 0.604 & 0.396 & 264 & 13.68 & 0.047 \\
\cline{2-8}
 & ABAABABBBBB & NM & 0.642 & 0.358 & 261 & 13.92 & 0.078 \\
\cline{2-8}
 & AAAABBBBBBB & NM & 0.717 & 0.283 & 255 & 14.43 & 0.134 \\
\cline{2-8}
 & AAABBBBBBBB & NM & 0.830 & 0.170 & 247 & 15.19 & 0.211 \\
\cline{2-8}
 & BBBBBBBBBAA & NM & 0.849 & 0.151 & 228 & 15.77 & 0.453 \\
\cline{2-8}
 & BBBBBBBBBBA & NM & 0.887 & 0.113 & 230 & 15.92 & 0.415 \\
\cline{2-8}
 & AABBBBBBBBB & NM & 0.943 & 0.057 & 240 & 15.94 & 0.284 \\
\cline{2-8}
 & ABBBBBBBBBB & NM & 0.981 & 0.019 & 238 & 16.20 & 0.306 \\
\cline{2-8}
 & BBBBBBBBBBB & NM & 1.000 & 0.000 & 238 & 16.31 & 0.308 \\
\cline{1-8} \cline{2-8}
\end{supertabular}

\end{center}
\endgroup

\pagebreak

\section{Spin polarized DFT calculations of simple TCP phases.}\label{SUP:SpinPolDFT}

\begin{figure}[htb!]
  \includegraphics[width=\textwidth]{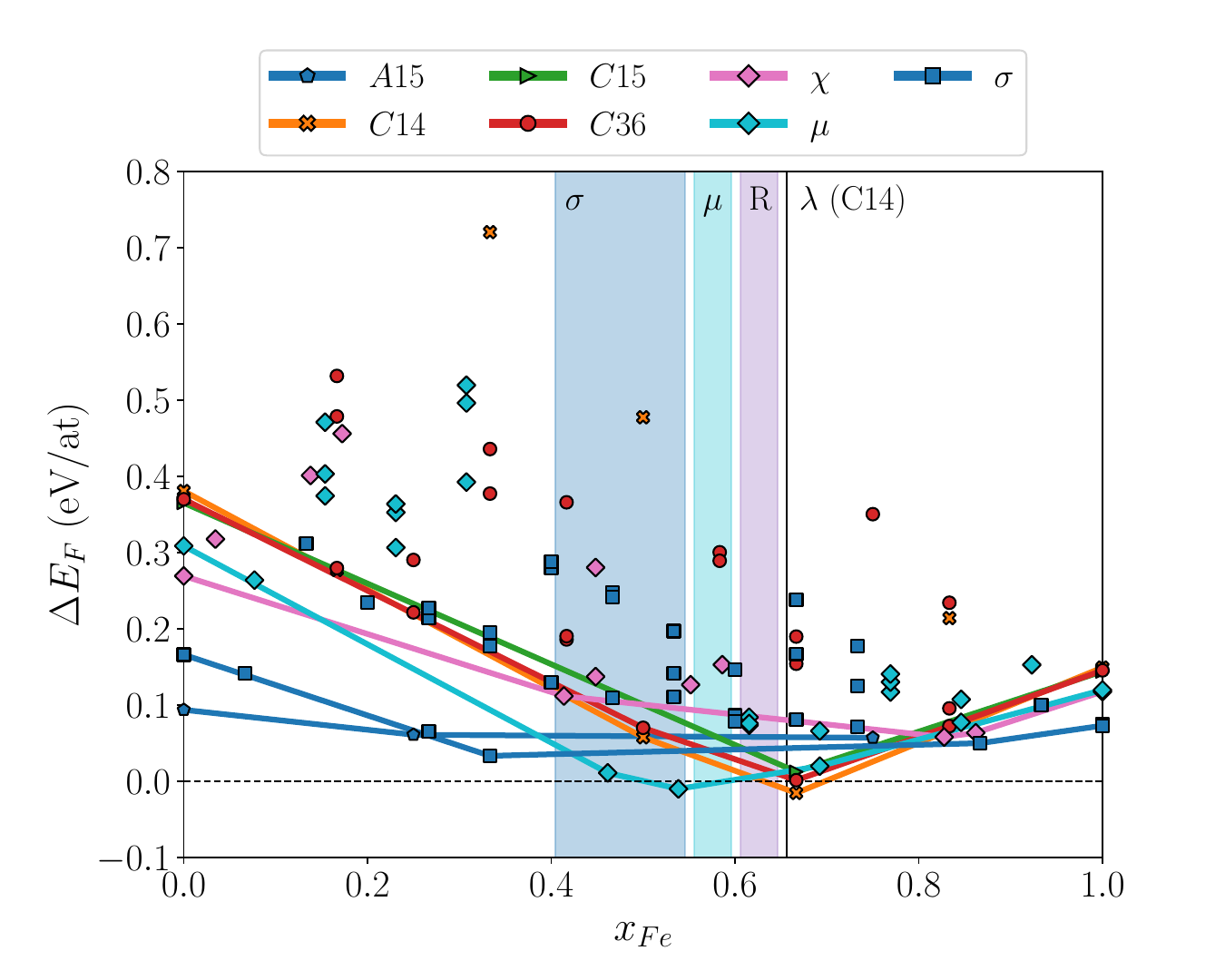}
  \caption{\protect\label{Fig:SpinPolDFT} 
Spin polarized  DFT calculations of the formation enthalpy of the simple TCP phases across the Fe--Mo composition range. 
  The vertical color bars indicate the experimentally observed stability ranges of the $\sigma$, $\mu$ and \textit{R} phases at finite temperatures. 
  The vertical black line indicates the  \textit{C}14 Fe\textsubscript{2}Mo line compound denoted as $\lambda$ in the phase diagram. Convex hull calculated from non-spin polarized calculations.
	}
\end{figure}

\pagebreak

\section{Comparison between estimated and fully optimized structures}\label{SUP:CompareGuessOpt}

\begin{figure}[h]
  \includegraphics[width=\textwidth]{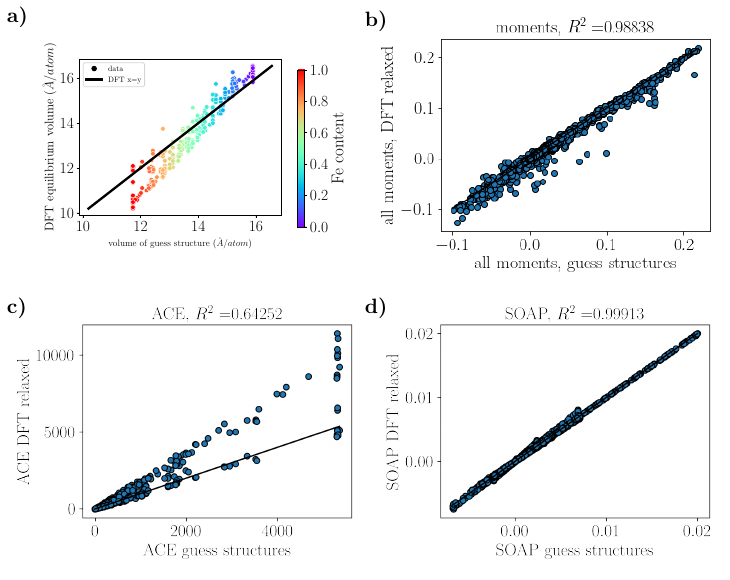}
  \caption{\protect\label{Fig:GuessOptCompare} Comparison of the  (a) volume per atom computed by DFT and by linear interpolation of the atomic volumes of the elements and respective descriptors computed with (b) BOP,   (c) ACE and (d) SOAP for the estimated and fully relaxed structures.
  }

\end{figure}

\autoref{Fig:GuessOptCompare} shows a comparison of the volumes of the simple Fe--Mo TCP phases obtained by DFT and the estimate using a linear interpolation of the volumes of the elements. In addition, the ACE, BOP and SOAP descriptors computed with these respective volumes show a good approximation to the  descriptors of the  equilibrium structure. 

%

\clearpage
\section{Distribution of WS in coordination polyhedra}

The interpretation of the spread in the formation energies for structures with the same composition requires the inspection of the atomic sites and the chemical nature of its coordination polyhedra. 
With this goal, we present here the collection of the multiplicity of the WS in the coordination polyhedra of each inequivalent WS of each phase. 
Such relations are ofthen found for simple TCP phases like $\sigma$ but there is no such compilation in the existing literature for all the TCPs.

\begin{minipage}[m][][b]{0.45\textwidth}
  \centering
  \captionof{table}{WS in CP of $A$15 phase.}
  \begin{tabular}{|l|c|c|c|}
\hline
 & 2$a$ & 6$c$ & total \\
\hline
2$a$ & 0 & 12 & 12 \\
6$c$ & 4 & 10 & 14 \\
\hline
\end{tabular}

\end{minipage}
\begin{minipage}[m][][b]{0.45\textwidth}
  \centering
  \captionof{table}{WS in CP of $C$15 phase.}
  \begin{tabular}{|l|c|c|c|c}
\hline
 & 8$a$ & 16$d$ & total \\
\hline
8$a$ & 4 & 12 & 16 \\
16$d$ & 6 & 6 & 12 \\
\hline
\end{tabular}

\end{minipage}

\begin{minipage}[m][][b]{0.45\textwidth}
  \centering
  \captionof{table}{WS in CP of $C$14 phase.}
  \begin{tabular}{|l|c|c|c|c|}
\hline
 & 4$f$ & 2$a$ & 6$h$ & total \\
\hline
4$f$ & 4 & 3 & 9 & 16 \\
2$a$ & 6 & 0 & 6 & 12 \\
6$h$ & 6 & 2 & 4 & 12 \\
\hline
\end{tabular}

\end{minipage} 
\begin{minipage}[m][][b]{0.45\textwidth}
  \centering
  \captionof{table}{WS in CP of $C$36 phase.}
  \begin{tabular}{|l|c|c|c|c|c|c|}
\hline
 & 4$e$ & 4$f_1$ & 6$g$ & 6$h$ & 4$f_2$ & total \\
\hline
4$e$ & 1 & 3 & 6 & 3 & 3 & 16 \\
4$f_1$ & 3 & 1 & 3 & 6 & 3 & 16 \\
6$g$ & 4 & 2 & 4 & 0 & 2 & 12 \\
6$h$ & 2 & 4 & 0 & 4 & 2 & 12 \\
4$f_2$ & 3 & 3 & 3 & 3 & 0 & 12 \\
\hline
\end{tabular}

\end{minipage}

\begin{minipage}[m][][b]{0.45\textwidth}
  \centering
  \captionof{table}{WS in CP of $\chi$ phase.}
  \begin{tabular}[t]{|l|c|c|c|c|c|}
\hline
 & 2$a$ & 8$c$ & 24$g_1$ & 24$g_2$ & total \\
\hline
2$a$ & 0 & 4 & 0 & 12 & 16 \\
8$c$ & 1 & 0 & 6 & 9 & 16 \\
24$g_1$ & 0 & 2 & 6 & 5 & 13 \\
24$g_2$ & 1 & 3 & 5 & 3 & 12 \\
\hline
\end{tabular}

\end{minipage}
\begin{minipage}[m][][b]{0.45\textwidth}
  \centering
  \captionof{table}{WS in CP of $\mu$ phase.}
  \begin{tabular}{|l|c|c|c|c|c|}
\hline
 & 1$a$ & 6$h$ & 2$c_1$ & 2$c_2$ & total \\
\hline
1$a$ & 0 & 6 & 0 & 6 & 12 \\
6$h$ & 1 & 4 & 2 & 5 & 12 \\
2$c_1$ & 0 & 6 & 3 & 6 & 15 \\
2$c_2$ & 3 & 9 & 0 & 4 & 16 \\
\hline
\end{tabular}

\end{minipage}

\begin{minipage}[m][][b]{0.45\textwidth}
  \centering
  \captionof{table}{WS in CP of $\sigma$ phase.}
  \begin{tabular}[t]{|l|c|c|c|c|c|c|}
\hline
 & 2$a$ & 4$f$ & 8$i_1$ & 8$i_2$ & 8$j$ & total \\
\hline
2$a$ & 0 & 4 & 0 & 4 & 4 & 12 \\
4$f$ & 2 & 1 & 2 & 4 & 6 & 15 \\
8$i_1$ & 0 & 1 & 5 & 4 & 4 & 14 \\
8$i_2$ & 1 & 2 & 4 & 1 & 4 & 12 \\
8$j$ & 1 & 3 & 4 & 4 & 2 & 14 \\
\hline
\end{tabular}

\end{minipage}
\begin{minipage}[m][][b]{0.45\textwidth}
  \centering
  \captionof{table}{WS in CP of $R$ phase.}
  \begin{tabular}{|l|c|c|c|c|c|c|c|c|c|c|c|c|}
\hline
 & ~~$b$~~ & 2$c_1$ & 2$c_2$ & 6$f_1$ & 6$f_2$ & 6$f_3$ & 6$f_4$ & 6$f_5$ & 6$f_6$ & 6$f_7$ & 6$f_8$ & total \\
\hline
$b$ & 0 & 0 & 0 & 0 & 0 & 1 & 6 & 5 & 0 & 0 & 0 & 12 \\
2$c_1$ & 0 & 0 & 0 & 1 & 3 & 3 & 2 & 1 & 2 & 0 & 0 & 12 \\
2$c_2$ & 0 & 0 & 2 & 3 & 2 & 0 & 0 & 0 & 1 & 6 & 2 & 16 \\
6$f_1$ & 0 & 0 & 1 & 0 & 1 & 2 & 0 & 2 & 1 & 2 & 3 & 12 \\
6$f_2$ & 0 & 1 & 1 & 2 & 1 & 2 & 0 & 1 & 1 & 0 & 3 & 12 \\
6$f_3$ & 0 & 1 & 0 & 2 & 1 & 0 & 1 & 2 & 2 & 1 & 2 & 12 \\
6$f_4$ & 1 & 1 & 0 & 1 & 0 & 2 & 2 & 3 & 2 & 1 & 1 & 14 \\
6$f_5$ & 1 & 0 & 1 & 1 & 1 & 1 & 3 & 2 & 2 & 0 & 2 & 14 \\
6$f_6$ & 0 & 1 & 1 & 0 & 1 & 3 & 3 & 2 & 2 & 1 & 1 & 15 \\
6$f_7$ & 0 & 0 & 2 & 2 & 1 & 1 & 0 & 0 & 2 & 2 & 2 & 12 \\
6$f_8$ & 0 & 0 & 2 & 3 & 2 & 3 & 1 & 1 & 1 & 2 & 1 & 16 \\
\hline
\end{tabular}

\end{minipage}

\begin{center}
\begin{minipage}[m][][b]{0.45\textwidth}
  \centering
  \captionof{table}{WS in CP of $P$ phase.}
  \begin{tabular}{|l|c|c|c|c|c|c|c|c|c|c|c|c|c|}
 \hline
 & 4$c_1$ & 4$c_2$ & 4$c_3$ & 4$c_4$ & 4$c_5$ & 4$c_6$ & 4$c_7$ & 4$c_8$ & 4$c_9$ & 4$c_{10}$ & 8$d_1$ & 8$d_2$ & total \\
 \hline
4$c_1$ & 0 & 1 & 1 & 0 & 2 & 3 & 0 & 1 & 0 & 0 & 2 & 2 & 12 \\
4$c_2$ & 1 & 0 & 1 & 2 & 2 & 0 & 1 & 0 & 1 & 0 & 0 & 4 & 12 \\
4$c_3$ & 1 & 1 & 0 & 1 & 2 & 0 & 0 & 0 & 2 & 1 & 2 & 2 & 12 \\
4$c_4$ & 0 & 2 & 1 & 0 & 1 & 0 & 2 & 1 & 2 & 1 & 0 & 4 & 14 \\
4$c_5$ & 2 & 2 & 2 & 1 & 0 & 1 & 0 & 0 & 1 & 0 & 2 & 4 & 15 \\
4$c_6$ & 3 & 0 & 0 & 0 & 1 & 2 & 1 & 3 & 0 & 0 & 4 & 2 & 16 \\
4$c_7$ & 0 & 1 & 0 & 2 & 0 & 1 & 0 & 2 & 1 & 3 & 2 & 2 & 14 \\
4$c_8$ & 1 & 0 & 0 & 1 & 0 & 3 & 2 & 0 & 0 & 1 & 2 & 2 & 12 \\
4$c_9$ & 0 & 1 & 2 & 2 & 1 & 0 & 1 & 0 & 0 & 3 & 2 & 2 & 14 \\
4$c_{10}$ & 0 & 0 & 1 & 1 & 0 & 0 & 3 & 1 & 3 & 2 & 4 & 0 & 15 \\
8$d_1$ & 1 & 0 & 1 & 0 & 1 & 2 & 1 & 1 & 1 & 2 & 2 & 0 & 12 \\
8$d_2$ & 1 & 2 & 1 & 2 & 2 & 1 & 1 & 1 & 1 & 0 & 0 & 2 & 14 \\
 \hline
\end{tabular}

\end{minipage}
\end{center}

\begin{center}
\begin{minipage}[m][][b]{0.45\textwidth}
  \centering
  \captionof{table}{WS in CP of $M$ phase.}
  \begin{tabular}{|l|c|c|c|c|c|c|c|c|c|c|c|c|}
 \hline
 & 4$c_6$ & 4$c_7$ & 4$c_8$ & 4$c_9$ & 8$d_1$ & 8$d_2$ & 4$c_1$ & 4$c_2$ & 4$c_3$ & 4$c_4$ & 4$c_5$ & total \\
\hline
4$c_1$ & 0 & 0 & 2 & 2 & 2 & 1 & 0 & 3 & 2 & 1 & 1 & 14 \\
4$c_2$ & 0 & 1 & 1 & 0 & 2 & 5 & 3 & 2 & 1 & 0 & 0 & 15 \\
4$c_3$ & 1 & 2 & 1 & 1 & 3 & 2 & 2 & 1 & 0 & 0 & 2 & 15 \\
4$c_4$ & 3 & 0 & 3 & 2 & 2 & 2 & 1 & 0 & 0 & 2 & 1 & 16 \\
4$c_5$ & 2 & 2 & 0 & 3 & 1 & 4 & 1 & 0 & 2 & 1 & 0 & 16 \\
4$c_6$ & 0 & 1 & 1 & 0 & 4 & 0 & 0 & 0 & 1 & 3 & 2 & 12 \\
4$c_7$ & 1 & 0 & 0 & 1 & 3 & 2 & 0 & 1 & 2 & 0 & 2 & 12 \\
4$c_8$ & 1 & 0 & 0 & 0 & 2 & 2 & 2 & 1 & 1 & 3 & 0 & 12 \\
4$c_9$ & 1 & 1 & 1 & 2 & 0 & 1 & 1 & 0 & 2 & 1 & 2 & 12 \\
8$d_1$ & 1 & 1 & 1 & 0 & 2 & 1 & 1 & 0 & 2 & 1 & 2 & 12 \\
8$d_2$ & 1 & 1 & 1 & 0 & 0 & 3 & 1 & 2 & 0 & 2 & 1 & 12 \\
\hline
\end{tabular}

\end{minipage}
\end{center}

\begin{center}
\begin{minipage}[m][][b]{0.45\textwidth}
  \centering
  \captionof{table}{WS in CP of $\delta$ phase.}
  \begin{tabular}{|l|c|c|c|c|c|c|c|c|c|c|c|c|c|c|c|}
  \hline
 & 4$c_1$ & 4$c_2$ & 4$c_3$ & 4$c_4$ & 4$c_5$ & 4$c_6$ & 4$c_7$ & 4$c_8$ & 4$c_9$ & 4$c_{10}$ & 4$c_{11}$ & 4$c_{12}$ & 4$c_{13}$ & 4$c_{14}$ & total \\
 \hline
4$c_1$ & 0 & 1 & 2 & 1 & 1 & 1 & 0 & 2 & 1 & 1 & 1 & 2 & 0 & 2 & 15 \\
4$c_2$ & 1 & 0 & 1 & 1 & 2 & 0 & 2 & 2 & 1 & 1 & 2 & 0 & 1 & 1 & 15 \\
4$c_3$ & 2 & 1 & 0 & 2 & 2 & 1 & 1 & 1 & 0 & 1 & 1 & 2 & 1 & 1 & 16 \\
4$c_4$ & 1 & 1 & 2 & 0 & 1 & 1 & 1 & 1 & 1 & 1 & 0 & 1 & 1 & 0 & 12 \\
4$c_5$ & 1 & 2 & 2 & 1 & 0 & 0 & 1 & 1 & 0 & 1 & 0 & 1 & 1 & 1 & 12 \\
4$c_6$ & 1 & 0 & 1 & 1 & 0 & 0 & 1 & 0 & 1 & 1 & 1 & 2 & 2 & 1 & 12 \\
4$c_7$ & 0 & 2 & 1 & 1 & 1 & 1 & 0 & 1 & 1 & 1 & 1 & 0 & 2 & 0 & 12 \\
4$c_8$ & 2 & 2 & 1 & 1 & 1 & 0 & 1 & 0 & 1 & 1 & 1 & 0 & 0 & 1 & 12 \\
4$c_9$ & 1 & 1 & 0 & 1 & 0 & 1 & 1 & 1 & 0 & 1 & 2 & 1 & 1 & 1 & 12 \\
4$c_{10}$ & 1 & 1 & 1 & 1 & 1 & 1 & 1 & 1 & 1 & 0 & 1 & 1 & 2 & 1 & 14 \\
4$c_{11}$ & 1 & 2 & 1 & 0 & 0 & 1 & 1 & 1 & 2 & 1 & 0 & 1 & 1 & 2 & 14 \\
4$c_{12}$ & 2 & 0 & 2 & 1 & 1 & 2 & 0 & 0 & 1 & 1 & 1 & 0 & 1 & 2 & 14 \\
4$c_{13}$ & 0 & 1 & 1 & 1 & 1 & 2 & 2 & 0 & 1 & 2 & 1 & 1 & 0 & 1 & 14 \\
4$c_{14}$ & 2 & 1 & 1 & 0 & 1 & 1 & 0 & 1 & 1 & 1 & 2 & 2 & 1 & 0 & 14 \\
\hline
\end{tabular}

\end{minipage}
\end{center}


\pagebreak

\section{Recursive feature addition with cross validation}\label{SECT_RCFSCV}

\begin{figure}[h!]
  \includegraphics[width=\textwidth]{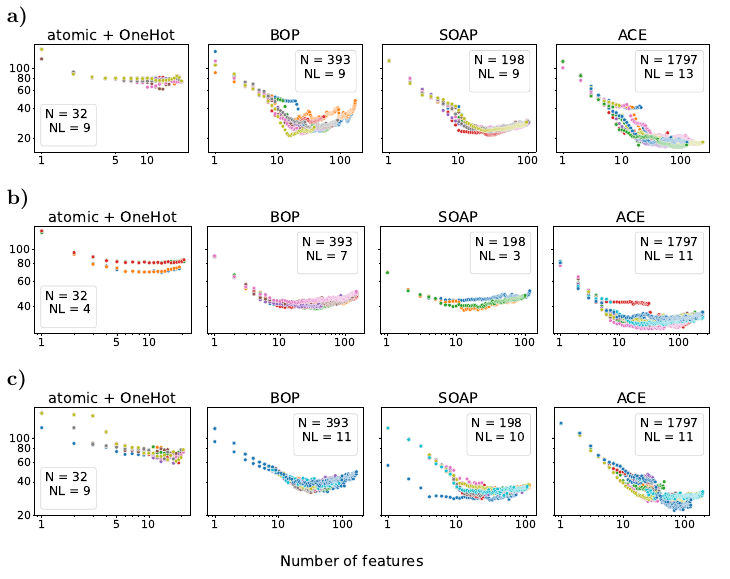}
  \caption{\protect\label{Fig:FeatureSelection}
    Feature selection curves. 
    Performance of the regression algorythm as a function of the selected number of features, measured from the test root mean squared error (y axis) on the test set. 
    All feature sets are inputs for (a) Kernel Ridge, (b) Random Forest and (c) Multi Layer Perceptron regressor.
    The labels inside the panels indicate the total number of features (N) and the total number of selection curves (NL).
    Each color indicate a separate recursive feature selection loop.
  }
\end{figure}

We present in \autoref{Fig:FeatureSelection} the learning curves for all the regression algorithms during the recursive feature addition process as explained in the Methods section. 
In all cases, the ACE features are less sensitive to overfitting, while BOP features are the most sensitive. 
The feature addition process was stopped when the test error increased over 10\% of the error at the optimal number of features.

\clearpage

\section{Train and test errors of the different feature/regression combinations}

\begin{center}

\tablecaption{\protect\label{TAB:ML_Results} 
Detail of the test error of the optimal ML models built from voting regressors for the individual optimal models from \autoref{Fig:FeatureSelection}. 
  This information is also plotted in \autoref{Fig:comparison_feature_sets}.
}

\tablehead{
\toprule[1.4pt]
 & \multicolumn{2}{c|}{~~~Kernel Ridge~~~} & \multicolumn{2}{c|}{~~~Random Forest~~~} & \multicolumn{2}{c}{~~~~~~MLP~~~~~~} \\
 & test & train & test & train & test & train \\
}
\begin{supertabular}{l|c|c|c|c|c|c}
\toprule
ACE & 0.016 & 0.007 & 0.029 & 0.020 & 0.028 & 0.029 \\
SOAP & 0.018 & 0.007 & 0.032 & 0.022 & 0.032 & 0.031 \\
ACE UCavg & 0.018 & 0.009 & 0.038 & 0.020 & 0.029 & 0.029 \\
BOP & 0.021 & 0.009 & 0.035 & 0.019 & 0.031 & 0.030 \\
SOAP UCavg & 0.022 & 0.014 & 0.032 & 0.019 & 0.032 & 0.034 \\
BOP UCavg & 0.037 & 0.023 & 0.046 & 0.022 & 0.038 & 0.041 \\
dataset UCavg & 0.071 & 0.078 & 0.121 & 0.037 & 0.108 & 0.087 \\
dataset & 0.075 & 0.057 & 0.060 & 0.026 & 0.064 & 0.055 \\
Canonical BOP CNavg ~~~ & 0.093 & 0.083 & 0.090 & 0.040 & 0.089 & 0.089 \\
Canonical BOP UCavg ~~~ & 0.113 & 0.089 & 0.136 & 0.074 & 0.109 & 0.103 \\
atomic & 0.157 & 0.146 & 0.150 & 0.104 & 0.146 & 0.146 \\
\hline
\end{supertabular}

\end{center}

\section{Formation energy of \textit{R}-phase configurations on convex hull}{\label{SUP:ConvexHullAnalysis}}


\begin{center}
  \tablecaption{\protect\label{TAB:CHULL_SAMPLES} 
    Formation energy of \textit{R}-phase configurations on convex hull. The sublattice occupation of the different configurations are given in terms of A/B for Fe/Mo. 
    Values in brackets indicate that the particular sublattice occupation is not predicted to be on the convex hull by the particular model.
	}
\tablehead{
  \toprule[1.4pt]
  {}
 & $x_{Fe}$ & $x_{Mo}$ & BOP & ACE & SOAP & DFT \\
  Configuration &  &  &  &  &  &  \\
  \toprule
}
\begin{supertabular}{l|c|c|c|c|c|c}
R-AAAAAAAAAAA & 1.000 & 0.000 & 0.165 & 0.168 & 0.168 & 0.166 \\
R-BAAAAAAAABA & 0.943 & 0.057 & [0.128] & 0.124 & [0.135] & [0.136] \\
R-AAAAAAAAAAB & 0.887 & 0.113 & 0.083 & 0.086 & 0.084 & 0.082 \\
R-AAAAAAAAABB & 0.849 & 0.151 & 0.066 & 0.064 & [0.066] & [0.063] \\
R-AAAAAAAABBB & 0.736 & 0.264 & 0.026 & [0.027] & [0.029] & [0.029] \\
R-AAAAAABAABB & 0.736 & 0.264 & [0.028] & [0.031] & 0.008 & 0.007 \\
R-BAAAAAAABBB & 0.717 & 0.283 & [0.024] & 0.013 & [0.028] & [0.022] \\
R-AAAAAAABBBB & 0.623 & 0.377 & -0.013 & -0.021 & -0.023 & -0.023 \\
R-AAAAAABBBBB & 0.509 & 0.491 & -0.032 & -0.037 & -0.042 & -0.043 \\
R-AAABAABBBBB & 0.396 & 0.604 & [0.040] & 0.026 & 0.023 & 0.023 \\
R-ABABAABBBBB & 0.358 & 0.642 & 0.057 & [0.065] & [0.053] & [0.059] \\
R-BBBBBBBBBBB & 0.000 & 1.000 & 0.309 & 0.307 & 0.302 & 0.308 \\
\hline
\end{supertabular}

\end{center}

%

\bibliographystyle{naturemag}
\bibliography{supplementary.bib}